\documentclass{emulateapj}          
\usepackage{natbib}
\usepackage{graphicx}
\usepackage{lscape}
\usepackage{amsmath}

\begin{document}                          

\title{Star Formation in Isolated Disk Galaxies. \\
II. Schmidt Laws and Efficiency of Gravitational Collapse}

\author{Yuexing Li\altaffilmark{1,2,4}, Mordecai-Mark Mac Low\altaffilmark{1,2}
and Ralf S. Klessen\altaffilmark{3}} 

\affil{$^{1}$Department of Astronomy, Columbia University, New York,
NY 10027, USA}
\affil{$^{2}$Department of Astrophysics, American Museum of Natural
History, 79th Street at Central Park West, New York, NY 10024-5192, USA}
\affil{$^{3}$Astrophysikalisches Institut Potsdam, An der Sternwarte
16, D-14482 Potsdam, Germany} 
\affil{$^4$ Current address: Harvard-Smithsonian Center for
  Astrophysics, 60 Garden Street, Cambridge, MA, 02138}
\email{yxli@cfa.harvard.edu, mordecai@amnh.org, rklessen@aip.de} 

\begin{abstract}
  We model gravitational instability in a wide range of isolated disk
  galaxies, using GADGET, a three-dimensional, smoothed particle hydrodynamics  
  code. The model galaxies include a dark matter halo and a disk of stars
  and isothermal gas. Absorbing sink particles are used to directly
  measure the mass of gravitationally collapsing gas. Below the
  density at which they are inserted, the collapsing gas is fully
  resolved. We make the assumption that stars and molecular gas form within the
  sink particle once it is created, and that the star formation rate is the
  gravitational collapse rate times a constant efficiency factor. In our
  models, the derived star formation rate declines exponentially with time,
  and radial profiles of atomic and molecular gas and star formation rate
  reproduce observed behavior.  We derive from our models and discuss both the
  global and local Schmidt laws for star formation: power-law relations
  between surface densities of gas and star formation rate. The global
  Schmidt law observed in disk galaxies is quantitatively reproduced by our
  models. We find that the surface density of star formation rate directly
  correlates with the strength of local gravitational instability.
  The local Schmidt laws of individual galaxies in our models show
  clear evidence of star formation thresholds. The variations in both
  the slope and the normalization of the local Schmidt laws cover
  the observed range. The averaged values agree well with the observed
  average, and with the global law. Our results suggest that the non-linear
  development of gravitational instability determines the local and global
  Schmidt laws, and the star formation thresholds. We derive from our models
  the quantitative dependence of the global star formation efficiency on the
  initial gravitational instability of galaxies. The more unstable a galaxy
  is, the quicker and more efficiently its gas collapses gravitationally and  
  forms stars.
\end{abstract}

\keywords{galaxy: evolution --- galaxy: spiral --- galaxy: kinematics
  and dynamics --- galaxy: ISM --- galaxy: star clusters --- stars:
  formation}

\section{INTRODUCTION}
Stars form at widely varying rates in different disk galaxies
\citep{kennicutt98a}. However, they appear to follow two simple empirical
laws. The first is the correlation between the star formation rate (SFR)
density and the gas density, the ``Schmidt law'' as first introduced by
\citet{schmidt59}: 
\begin{equation}
\label{eq_schmidt}
\Sigma_{\rm SFR} = A \ \Sigma_{\rm gas}^{\rm N} 
\end{equation} 
where $\Sigma_{\rm SFR}$ and $\Sigma_{\rm gas}$ are the surface densities of
SFR and gas, respectively. 

When $\Sigma_{\rm gas}$ and $\Sigma_{\rm SFR}$ are averaged over the
entire star forming region of a galaxy, they give rise to a global
Schmidt law.  \citet{k98} found a universal global star formation law
in a large sample that includes 61 normal spiral galaxies that have
$\rm H_{\alpha}$, H$_{\rm I}$ and CO measurements and 36
infrared-selected starburst galaxies. The observations show that both
the slope $N \sim$~1.3--1.5 and the normalization $A$ appear to be
remarkably consistent from galaxy to galaxy. There are some variations,
though. For example, \citet{wong02} reported $N \sim$~1.1--1.7 for a sample of
seven molecule-rich spiral galaxies, depending on the correction of the
observed $\rm H_{\alpha}$ emission for extinction in deriving the star
formation rate. \citet{boissier03} examined 16 spiral galaxies with published
abundance gradients and found $N \sim$ 2.0. \citet{gao04a} surveyed HCN
luminosity, a tracer of dense molecular gas, from 65 infrared or CO-bright
galaxies including nearby normal spiral galaxies, luminous infrared galaxies,
and ultraluminous infrared galaxies. Based on this survey, \citet{gao04b}
suggested a shallower star formation law with a power-law index of 1.0 in
terms of dense molecular gas content.  

When $\Sigma_{\rm gas}$ and $\Sigma_{\rm SFR}$ are measured radially
within a galaxy, a local Schmidt law can be measured. \citet{wong02}
investigated the local Schmidt laws of individual galaxies in
their sample. They found similar correlations in these galaxies but
the normalizations and slopes vary from galaxy to galaxy, with $N
\sim$~1.2--2.1 for total gas, assuming that extinction depends
on gas column density (or, $N \sim$ 0.8--1.4 if extinction is
assumed constant). \citet{heyer04} reported that M33 has a much deeper
slope, $N \simeq 3.3$.

The second empirical law is the star formation threshold. Stars are
observed to form efficiently only above a critical gas surface density.
\citet{martin01} studied a sample of 32 nearby spiral galaxies with
well-measured $\rm {H_{\alpha}}$ and H$_{2}$ profiles, and
demonstrated clear surface-density thresholds in the star formation
laws in these galaxies. They found that the threshold gas density
(measured at the outer threshold radius where SFR drops sharply)
ranges from 0.7 to 40 $\rm {M_{\odot} \ pc^{-2}}$ among spiral
galaxies, and the threshold density for molecular gas is $\sim$5--10
$\rm {M_{\odot} \ pc^{-2}}$. However, they found that the ratio of gas
surface density at the threshold to the critical density for
\citet{toomre64} gravitational instability, $\alpha_Q =
\Sigma_{\rm gas}/\Sigma_{\rm crit}$, is remarkably uniform with
$\alpha_Q = 0.69 \pm 0.2$.  They assumed a constant velocity
dispersion of the gas, the effective sound speed, of $c_s =
6$~km~s$^{-1}$. Such a density threshold applies to normal disk
galaxies (e.g., \citealt{boissier03}), elliptical galaxies (e.g.,
\citealt{vader91}), low surface brightness galaxies
\citep{vanderhulst93}, and starburst galaxies \citep{elmegreen94a}.
However, there are a few exceptions, such as dwarf and irregular
galaxies (e.g., \citealt*{hunter98}). Furthermore, inefficient star formation
can be found well outside the threshold radius \citep{ferguson98}.

What is the origin of the Schmidt laws and the star formation thresholds? The
mechanisms that control star formation in galaxies, such as gravitational
instability, supersonic turbulence, magnetic fields, and rotational shear are
widely debated (\citealt{shu87, elmegreen02, larson03, mk04}). At least four
types of models are currently discussed. The first type emphasizes
self-gravity of the galactic disk (e.g., \citealt{quirk72, larson88,
  kennicutt89, elmegreen94a, k98}). In these models, the Schmidt laws do not
depend on the local star formation process, but are simply the results of
global gravitational collapse on a free-fall time. In the second type, the
global star formation rate scales with either the local dynamical time,
invoking cloud-cloud collisions (e.g. \citealt{wyse86, wyse87, silk97,
  tan00}), or the local orbital time of the galactic disk (e.g.\
\citealt{elmegreen97, hunter98}). A third type, which invokes hierarchical
star formation triggered by turbulence, has been proposed by
\citet{elmegreen02}. In this model, the Schmidt law is scale-free, and the
star formation rate depends on the probability distribution function (PDF) of
the gas density produced by galactic turbulence, which appears to be 
log-normal in simulations of turbulent molecular clouds and interstellar
medium (e.g., \citealt*{scalo98, passot98, ostriker99, klessen00, wada01,
  bp02, padoan02, li03, kravtsov03, maclow05}). Recently, \citet{krumholz05}
extended this analysis with additional assumptions such as the virialization
of the molecular clouds and star formation efficiency to derive the star
formation rate from the gas density PDFs. They successfully fitted the global
Schmidt law, but their theory still contains several free or poorly
constrained parameters, and does not address the observed variation in local 
Schmidt laws among galaxies. A fourth type appeals to the gas dynamics and
thermal state of the gas to determine the star formation behavior.
\citet{struck91} and citet{struck99}, for example, suggest that galactic disks
are in thermohydrodynamic equilibrium maintained by feedback from star
formation and countercirculating radial gas flows of warm and cold gas.

There is considerable debate on the star formation threshold as well.
\citet{martin01} suggest that the threshold density is determined by
the Toomre criterion \citep{toomre64} for gravitational instability.
\citet{hunter98} argued that the critical density for star formation
in dwarf galaxies depends on the rotational shear of the disk.
\citet{wong02} claimed no clear evidence for a link between
$\alpha_Q$ and star formation. Instead, they suggested that
$\alpha_Q$ is a measurement of gas fraction. \citet{boissier03}
found that the gravitational instability criterion has limited
application to their sample. Note all the models above are based on an
assumption of constant sound speed for the gas. \citet{schaye04}
proposed a thermal instability model for the threshold, in which the
velocity dispersion or effective temperature of the gas is not
constant, but drops from a warm (i.e.  10$^{4}$ K) to a cold phase
(below 10$^{3}$ K) at the threshold. He suggested that such a
transition is able to reproduce the observed threshold density.

While each of these models has more or less succeeded in explaining
the Schmidt laws or the star formation threshold, a more complete
picture of star formation on a galactic scale remains needed.
Meanwhile, observations of other properties related to star formation in
galaxies have provided more clues to the dominant mechanism that controls
global star formation. 

An analysis of the distribution of dust in a sample of 89 edge-on,
bulgeless disk galaxies by \citet*{dalcanton04} shows that dust lanes
are a generic feature of massive disks with $V_{\rm rot} >
120$~km~s$^{-1}$, but are absent in more slowly rotating galaxies with
lower mass.  These authors identify the $V_{\rm rot} =
120$~km~s$^{-1}$ transition with the onset of gravitational
instability in these galaxies, and suggest a link between the disk
instability and the formation of the dust lanes which trace star
formation. 

Color gradients in galaxies help trace their star formation history by
revealing the distribution of their stellar populations
\citep{searle73}. A comprehensive study of color gradients in 121
nearby disk galaxies by \citet{bell00} shows that the star formation
history of a galaxy is strongly correlated with the surface mass
density. Similar conclusions were drawn by \citet{kauffmann03} from a
sample of over $10^5$ galaxies from the Sloan Digital Sky Survey.
Recently, \citet{macarthur04} carried out a survey of 172
low-inclination galaxies spanning Hubble types S0--Irr to investigate
optical and near-IR color gradients.  These authors find strong
correlations in age and metallicity with Hubble type, rotational
velocity, total magnitude, and central surface brightness. Their
results show that early type, fast rotating, luminous, or high surface
brightness galaxies appear to be older and more metal-rich than their
late type, slow rotating, or low surface brightness counterparts,
suggesting an early and more rapid star formation history for the
early type galaxies.

These observations show that star formation in disks correlates well
with the properties of the galaxies such as rotational velocity,
velocity dispersion, and gas mass, all of which directly determine the
gravitational instability of the galactic disk. This suggests that, on
a galactic scale, gravitational instability controls star formation.

The nonlinear development of gravitational instability and its effect
on star formation on a galactic scale can be better understood through
numerical modeling. There have been many simulations of disk galaxies,
including isolated galaxies with various assumptions of the gas physics and
feedback effects (e.g., \citealt*{thacker00, wada01, noguchi01, barnes02,
  robertson04, li05a, okamoto05}), galaxy mergers (e.g., \citealt*{mihos94,
  barnes96, li04}), and galaxies in a cosmological context, with different
assumptions about the nature and distribution of dark matter (e.g.,
\citealt*{katz91, navarro91, katz92, steinmetz94, nfw95, sommer-larsen99,
  steinmetz99, springel00, sommer-larsen01, sommer-larsen03, springel03,
  governato04}). However, in most of these simulations gravitational collapse
and star formation are either not numerically resolved, or are followed with
empirical recipes tuned to reproduce the observations {\em a priori}.
There are only a handful of numerical studies that focus on the star formation
laws. Early three-dimensional smoothed particle hydrodynamics (SPH)
simulations of isolated barred galaxies were carried out by \citet*{friedli93,
  friedli94}, and \citet{friedli95}. In \citet{friedli93}, the secular
evolution of the isolated galaxies was followed by modeling of a two-component
(gas and stars) fluid, restricting the interaction between the two to purely
gravitational coupling. The simulations were improved later in
\citet{friedli95} by including star formation and radiative cooling.  These
authors found that their method to simulate star formation, based on Toomre's
criterion, naturally reproduces both the density threshold of $7 \ \rm
M_{\odot}\ pc^{-2}$ for star formation, and the global Schmidt law in
disk galaxies. They also found that the nuclear starburst is
associated with bar formation in the galactic center. \citet{gerritsen97}
included stellar feedback in similar two-component (gas and stars) simulations 
that yielded a Schmidt law with power-law index of $\sim$ 1.3. 
However, these simulations included only stars and gas, and no dark matter. 
More recently, \citet{kravtsov03} reproduced the global Schmidt law using
self-consistent cosmological simulations of high-redshift galaxy
formation. He argued that the global Schmidt law is a
manifestation of the overall density distribution of the interstellar
medium, and that the global star formation rate is determined by the
supersonic turbulence driven by gravitational instabilities on large
scales, with little contribution from stellar feedback. However, the
strength of gravitational instability was not directly measured in
this important work, so a direct connection could not be made between
instability and the Schmidt laws.

In order to investigate gravitational instability in disk galaxies and
consequent star formation, we model isolated galaxies with a wide
range of masses and gas fraction. In \citet[][hereafter Paper
I]{li05a} we have described the galaxy models and computational
methods, and discussed the star formation morphology associated with
gravitational instability. In that paper it was shown that the
nonlinear development of gravitational instability determines where
and when star formation takes place, and that the star formation
timescale $\tau_{\rm SF}$ depends exponentially on the initial Toomre
instability parameter for the combination of collisonless stars and
collisional gas in the disk $Q_{sg}$ derived by \citet{rafikov01}. Galaxies
with high initial mass or gas fraction have small $Q_{sg}$ and are more
unstable, forming stars quickly, while stable galaxies with $Q_{sg} > 1$
maintain quiescent star formation over a long time.

Paper~I emphasized that to form a stable disk and derive the correct
SFR from a numerical model, the gravitational collapse of the gas must
be fully resolved \citep{bate97,truelove97} up to the density where
gravitationally collapsing gas decouples from the flow.  If this is done,
stable disks with SFRs comparable to observed values can be derived from
models using an isothermal equation of state.  With insufficient resolution,
however, the disk tends to collapse to the center producing much higher SFRs,
as found by some previous work \citep[e.g.][]{robertson04}. 

We analyze the relation between the SFR and the gas density, both globally and
locally. In \S~\ref{sec_com} we briefly review our computational method,
galaxy models and parameters. In \S~\ref{sec_dist} we present the evolution of
the star formation rate and radial distributions of both gas and star
formation. We derive the global Schmidt law in \S~\ref{sec_global}, followed
by a parameter study and an exploration of alternative forms of the star
formation law. Local Schmidt laws are presented in \S~\ref{sec_local}.  In
\S~\ref{sec_sfe} we investigate the star formation efficiency. The assumptions
and limitations of the models are discussed in \S~\ref{sec_dis}.  Finally, we
summarize our work in \S~\ref{sec_sum}. Preliminary results on the global
Schmidt law and star formation thresholds were already presented by
\citet*{li05b}.  

\section{COMPUTATIONAL METHOD}
\label{sec_com}
We here summarize the algorithms, galaxy models, and numerical parameters
described in detail in Paper I.  We use the SPH code GADGET, v1.1
\citep*{springel01}, modified to include absorbing sink particles
\citep*{bate95} to directly measure the mass of gravitationally collapsing
gas. Paper I and \citet{jappsen05} give detailed descriptions of sink particle
implementation and interpretation. In short, a sink particle is created from
the gravitationally bound region at the stagnation point of a converging flow
where number density exceeds values of $n = 10^3$~cm$^{-3}$. It interacts
gravitationally and inherits the mass, and linear and angular momentum of the
gas. It accretes surrounding gas particles that pass within its accretion
radius and are gravitationally bound. 

Regions where sink particles form have pressures $P/k \sim 10^7 $~K~cm$^{-3}$
typical of massive star-forming regions.  We interpret the formation of sink
particles as representing the formation of molecular gas and stellar clusters.
Note that the only regions that reach these high pressures in our simulations
are dynamically collapsing. The measured mass of the collapsing gas is
insensitive to the value of the cutoff-density.  This is not an important free
parameter, unlike in the models of \citet{elmegreen02} and \citet{krumholz05}.

Our galaxy model consists of a dark matter halo, and a disk of stars and
isothermal gas.  The initial galaxy structure is based on the analytical work
by \citet*{mo98}, as implemented numerically by \citet{springel99,springel00}
and \citet*{springel05}. We characterize our models by the rotational velocity
$V_{200}$ at the virial radius $R_{200}$ where the density reaches 200 times
the cosmic average.  We have run models of galaxies with rotational velocity
$V_{200} = 50$--220~km~s$^{-1}$, with gas fractions of 20--90\% of the disk
mass for each velocity.

Observations of H$_{\rm I}$ in many spiral galaxies suggest that the
gas velocity dispersion has a range of $\sim$4--15~km~s$^{-1}$ (e.g.,
see review in \citealt*{dib05}). The dispersion $\sigma_{\rm H_{\rm I}}$
varies radially from $\sim 12$--15~km~s$^{-1}$ in their central regions to
$\sim$4--6~km~s$^{-1}$ in the outer parts (e.g., \citealt{vanderkruit82,
  dickey90, kamphuis93, rownd94, meurer96}). We therefore choose two sets of
effective sound speeds for the gas, $c_s = 6$~km~s$^{-1}$ (low
temperature models) as suggested by \citet{k98}, and $c_s = 15$ km
s$^{-1}$ (high temperature models). Table 1 lists the most important
model parameters. The Toomre criterion for gravitational instability
that couples stars and gas, $Q_{sg}$ is calculated following
\citet{rafikov01}, and the minimum value is derived using the
wavenumber $k$ of greatest instability and lowest $Q_{sg}$ at each
radius.

Models of gravitational collapse must satisfy three numerical
criteria: the Jeans resolution criterion (\citealt{bate97}, hereafter
BB97; \citealt{whitworth98}), the gravity-hydro balance criterion for
gravitational softening (BB97), and the equipartition criterion for
particle masses \citep{steinmetz97}. We set up our simulations to
satisfy the above three numerical criteria, with the computational
parameters listed in Table~1. We choose the particle number for each
model such that they not only satisfy the criteria, but also so that
all runs have at least $10^6$ total particles. The gas, halo and
stellar disk particles are distributed with number ratio $N_{\rm g}$ :
$N_{\rm h}$ : $N_{\rm d}$ = 5 : 3 : 2. The gravitational softening
lengths of the halo $h_{\rm h} = 0.4$~kpc and disk $h_{\rm d}=
0.1$~kpc, while that of the gas $h_{\rm g}$ is given in Table~1 for
each model.  The minimum spatial and mass resolutions in the gas are
given by $h_{\rm g}$ and twice the kernel mass ($\sim 80 m_g$). (Note that
we use $h_g$ here to denote the gravitational softening length instead of
$\epsilon$ as used in previous papers, to distinguish it from the
star formation efficiency used in later sections.) We adopt typical values for
the halo concentration parameter $c = 5$, spin parameter $\lambda = 0.05$, and
Hubble constant $H_0 = 70$ km s$^{-1}$ Mpc$^{-1}$ \citep{springel00}.

Resolution of the Jeans length is vital for simulations of
gravitational collapse \citep{truelove97, bate97}. Exactly how well
the Jeans length must be resolved remains a point of controversy,
however. \citet{truelove97} suggest that a Jeans mass must be resolved
with far more than the $N_k = 2$ smoothing kernels proposed by BB97.
In this work, we carried out a resolution study on low-$T$ models
G100-1 and G220-1, with three resolution levels having total particle
numbers $N_{\rm tot} = 10^5$ (R1), $8 \times 10^5$ (R8) and $6.4\times
10^6$ (R64), reaching $N_{\rm k}\approx$ 24. The particle numbers are
chosen such that the maximum spatial resolution increases by a factor
of two between each pair of runs. Paper I finds convergence to within
10\% of the global amount of mass accreted by sink particles between
the two highest resolutions, suggesting that the BB97 criterion is
sufficient for the problem of global collapse in galactic disks. In
this paper we also refer to the resolution study as applicable.

\section{STAR FORMATION AND GAS DISTRIBUTION}
\label{sec_dist}

Molecular clouds and stars form together in galaxies. 
Molecular hydrogen forms on dust grains in a time of \citep*{hollenbach71}:  
\begin{equation}
t_f \simeq (10^9 \mbox{ yr}) / n \label{tform}
\end{equation}
where $n$ is the number density of the gas.  The
absence of $\gtrsim 10$~Myr old stars in star-forming regions in the solar
neighborhood suggests that molecular cloud complexes must coalesce
rapidly and form stars quickly \citep*{bhv99, hartmann00}.
\citet{hartmann01} further suggested that the conditions needed for
molecular gas formation from atomic flows are similar to the
conditions needed for gravitational instability. Star formation can
therefore take place within a free-fall timescale once molecular
clouds are produced. Using a one-dimensional chemical model,
\citet{bergin04} showed rapid formation of molecular gas in 12--20~Myr
in shock-compressed regions. These results are confirmed by
\citet{glover05} using three-dimensional magnetohydrodynamics
simulations with chemistry of supersonic turbulence.  They show that
most of the atomic gas turns into H$_2$ in just a few megayears once
the average gas density rises above $\sim 100$~cm$^{-3}$.  This is because the
gas passes through turbulent density fluctuations of higher density where H$_2$
can form quickly.  By the time gas reaches the densities of $n =
1000$~cm$^{-3}$ where we replace it with sink particles, the molecular
hydrogen formation timescale $t_f \lesssim 1$~Myr (eq.~\ref{tform}).
Motivated by these results, we identify the high-density regions
formed by gravitational instability as giant molecular cloud complexes
and replace these regions by accreting sink particles.

We assume that a fraction of the molecular gas turns into stars
quickly.  CO observations by \citet{young96} and \citet{rownd99}
suggest that the {\em local} star formation efficiency (SFE) in
molecular clouds remains roughly constant. To quantify the SFR, we
therefore assume that individual sink particles form stars at a fixed local 
efficiency $\epsilon_{\ell}$.

\citet{k98} found a global SFE of $\epsilon_{\rm g} = 30$\% for
starburst galaxies, which \citet{wong02} found to be dominated by
molecular gas.  We take this to be a measure of the local SFE
$\epsilon_{\ell}$ in individual molecular clouds, since most gas in
these galaxies has already become molecular.  In our simulations, sink
particles represent high pressure ($P/k \approx 10^7$ K cm$^{-3}$),
massive star formation regions in galaxies, such as 30 Doradus in the
Large Magellanic Cloud (e.g., \citealt{walborn99}). We therefore adopt
a fixed local SFE of $\epsilon_{\ell} = 30$\% to convert the mass of
sink particles to stars, while making the simple approximation that
the remaining 70\% of the sink particle mass remains in molecular
form. This approach will be discussed in more detail in
\S~\ref{sec_sfe}.
 
\subsection{Evolution of Star Formation Rate}
\label{subsec_sfr}

\begin{figure}
\begin{center}
\includegraphics{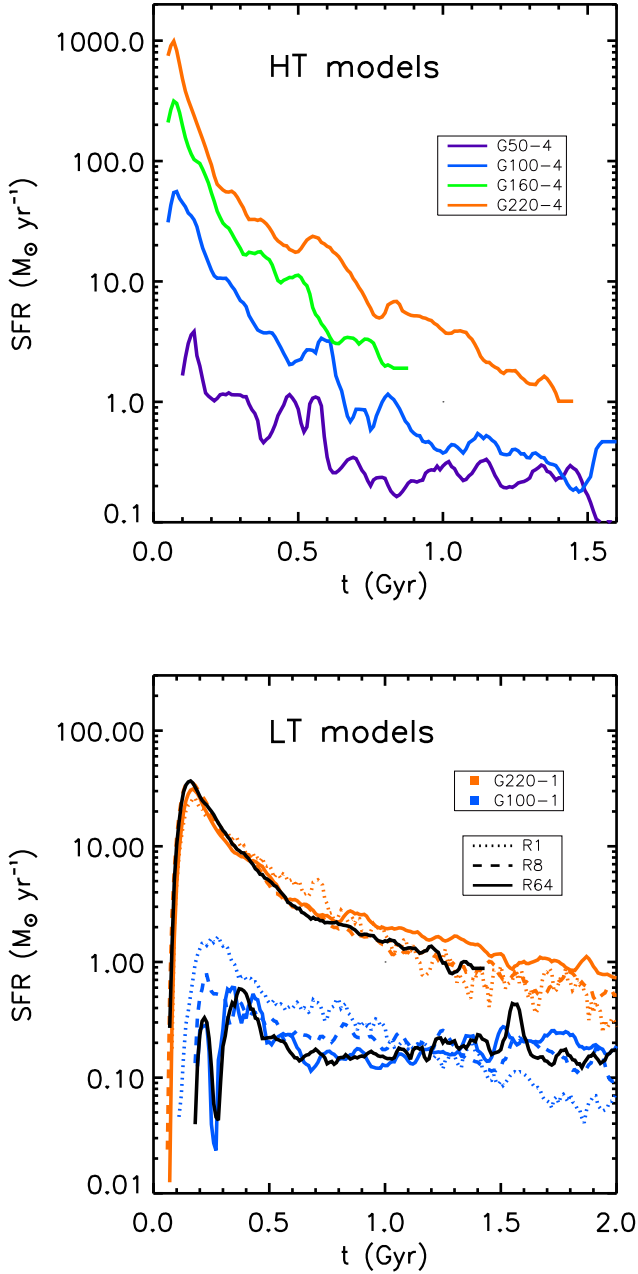}
\caption[Time Evolution of SFR]{\label{fig_sfr}  ({\em a}) Time
  evolution of the SFRs in selected high-$T$ models as given in the
  legend. ({\em b}) Resolution study of low-$T$ models G100-1 ({\em
    blue}) and G220-1 ({\em orange}) with resolutions of R1 ({\em
    dotted lines}), R8 ({\em dashed lines}) and R64 ({\em solid
    colored lines}), where the resolution levels are in units of
  $10^5$ total particles. The standard R10 models ({\em black}) are
  shown for comparison.}
\end{center} 
\end{figure}

Figure~\ref{fig_sfr} shows the time evolution of the SFRs 
of different models. The SFR is calculated as $\rm SFR = dM_{*}/dt \simeq
\epsilon_{\ell}\ \Delta\  M_{\rm sink}/\Delta t$, where $M_{*}$ and $M_{\rm
  sink}$ are the masses of the stars and sink particles, respectively. We choose
$\epsilon_{\ell}=30\%$ to be the local SFE within sink particles, and the
time interval $\Delta t = 50$ Myr.  Figure~\ref{fig_sfr}a shows the SFR
curves of selected high~$T$ models, while Figure~\ref{fig_sfr}b shows a
resolution study of SFR evolution. We find convergence to within 10\% of the
SFR between the two highest resolutions over periods of more than 2 Gyr,
suggesting that this result converges well under the BB97 criterion.

\begin{figure}
\begin{center}
\includegraphics{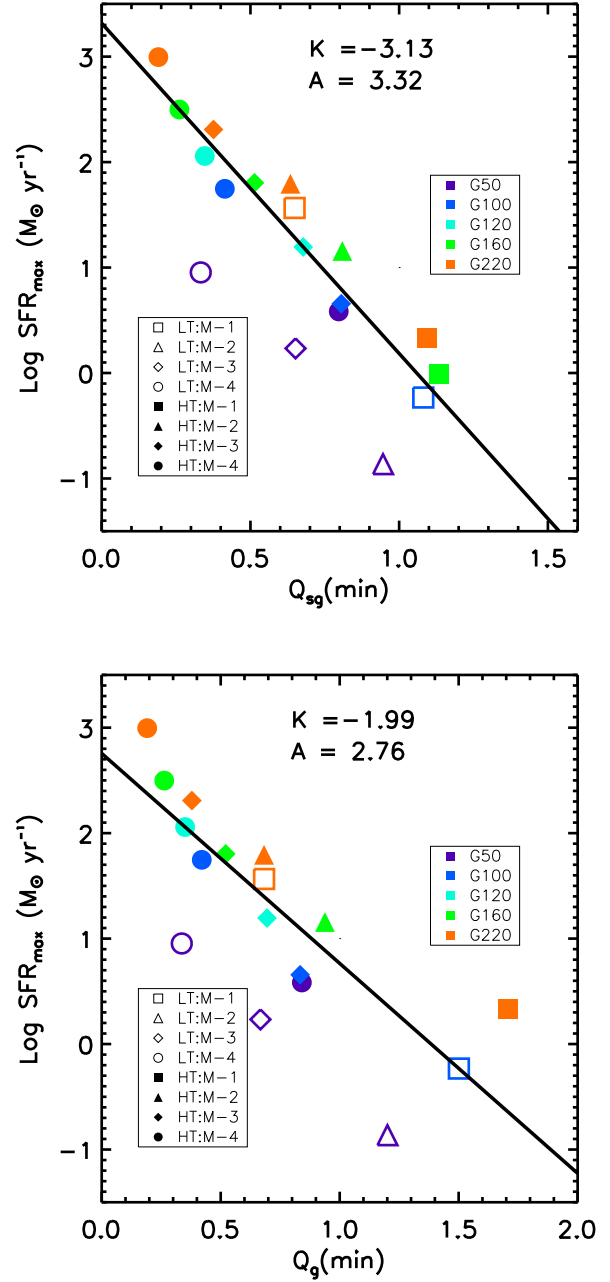}
\caption{\label{fig_sfr_max}Correlation between the maximum of the SFR
and the minimum radial values of the Toomre instability parameters
  ({\em a}) for stars and gas together $Q_{sg,{\rm min}}$  and ({\em b})
  for gas alone $Q_{g,{\rm min}}$. The solid lines are the least absolute
  deviation fits $\log {\rm SFR}_{\rm max} = K + A Q_{\rm min}$.}    
\end{center} 
\end{figure}

The star formation rates in Figure~\ref{fig_sfr} decline over time.
Many of the models have SFR~$ > 10 \rm M_{\odot}$~yr$^{-1}$,
corresponding to starburst galaxies.  Some small or gas poor models
such as G100-1, on the other hand, have SFR~$ < 1 \rm
M_{\odot}$~yr$^{-1}$. They maintain slow but steady star formation
over a long time and may represent quiescent normal galaxies. The
maximum SFR appears to depend quantitatively on the initial
instability of the disk as measured either by $Q_{sg, {\rm min}}$, the
minimum Toomre parameter for the combination of stars and gas in the
disk, or by the value for the gas only $Q_{g,{\rm min}}$.
Figure~\ref{fig_sfr_max} shows both correlations: $\log {\rm SFR}_{\rm
  max} \simeq 3.32 - 3.13Q_{sg,{\rm min}}$ and $\log {\rm SFR}_{\rm
  max} \simeq 2.76 - 1.99 Q_{g,{\rm min}}$.

The SFRs in most models shown in Figure~\ref{fig_sfr} appear to decline
exponentially. From Paper I, the accumulated mass of the stars formed in 
each galaxy can be fitted with an exponential function: 
\begin{equation}
\label{eq_mas}
\frac{M_{*}}{M_{\rm init}} =  M_0\ \left[1 - \exp\left({-t}/{\tau_{\rm
        SF}}\right)\right]\,,
\end{equation} 
where 
\begin{equation}
\label{eq_m0}
M_0 =  0.96\ \epsilon_{\ell}\ \{1 - 2.9\ \exp\left[{-1.7}/{Q_{sg,{\rm min}}}\right]\} \,, 
\end{equation}

\begin{equation}
\label{eq_tau}
\tau_{\rm SF} = \left(34 \pm 7\  \mbox{Myr}\right)\
\exp\left[Q_{sg,{\rm min}}/0.24\right], 
\end{equation}
$M_{\rm init}$ is the initial total gas mass, and $\tau_{\rm
  SF}$ is the star formation timescale. The star formation rate can then be
  rewritten in the following form: 
\begin{equation}
\label{eq_sfr}
\mbox{SFR} = \frac{dM_{*}}{dt} \propto \frac{1}{\tau_{\rm SF}} {\exp\left({-t}/{\tau_{\rm
        SF}}\right)} \,.
\end{equation} 
A similar exponential form is also reported by \citet{macarthur04}, as first suggested by
\citet{larson74} and \citet{tinsley78}.

\citet{sandage86} studied the star formation rate of different types of galaxies in
the Local Group and proposed an alternative  form for the star formation
history, as explicitly formulated by \citet{macarthur04}:
\begin{equation}
\label{eq_sfr_san}
\mbox{SFR} \propto \frac{t}{\tau_{\rm SF}^2} \exp\left({-t^2}/{\tau_{\rm
        SF}^2}\right)
\end{equation} 
 
\begin{figure}
\begin{center}
\includegraphics{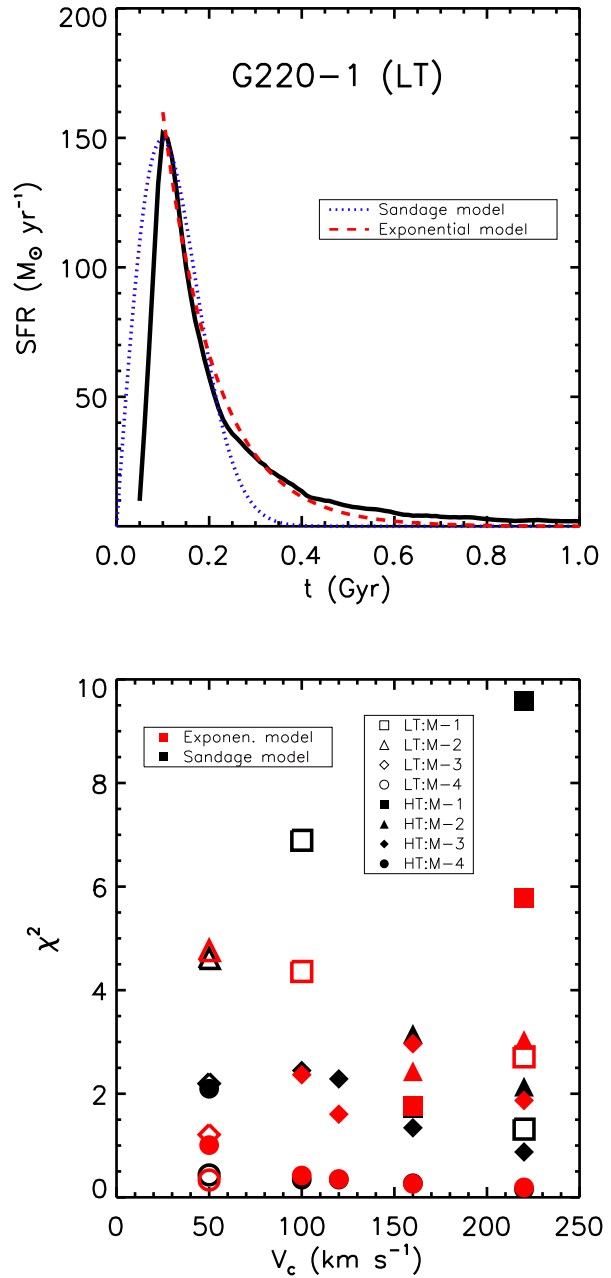}
\caption[Fitting of the SFR -- t Curves]{\label{fig_sfr_fit} 
  ({\em a}) Example fit of the SFR evolution curve with the
  exponential form (eq.~\ref{eq_sfr}) and the \citet{sandage86} form
  (eq.~\ref{eq_sfr_san}) for model G220-1 (low-$T$) as an example.
  ({\em b}) The relative goodness of fit $\chi^2$ (eq.~\ref{chi2}) for
  all models for exponential ({\em red}) and \citet{sandage86} ({\em
    black}) forms.  }
\end{center} 
\end{figure}

Figure~\ref{fig_sfr_fit}a shows an example of model G220-1 (low-$T$) fitted with
these two formula. Both formulae appear quite similar at intermediate times.
The Sandage model captures the initial rise in star formation better, but the
exponential form follows the late time behavior of our models more closely.
As we only include stellar feedback implicitly by maintaining constant gas
sound speed, we must be somewhat cautious about our interpretation of the late
time results. In order to compare the fits, we define a parameter for relative
goodness of the fit
\begin{equation} \label{chi2}
\chi^2 = \sum \left([y_s - y_f]/y_{m}\right)^2
\end{equation}
where $y_s$ is the SFR from the simulation, $y_{m}$ is the maximum of SFR, and
$y_f$ is the model function from equation (\ref{eq_sfr}) or
(\ref{eq_sfr_san}). Note that since we do not take into account the 
uncertainty of each point, the absolute value of $\chi^2$ has no
meaning. We only compare the relative $\chi^2$ values in
Figure~\ref{fig_sfr_fit}{\em b}. Both formulae fit equally well to many 
models, especially to those with high gas fractions that form a lot of stars
early on. But for some models such as G100-1 (low-$T$) and G220-1 (high-$T$),
the exponential function seems to fit noticeably better.  Therefore we use
the exponential form in the rest of the paper.

This analysis implies that the star formation history depends quantitatively on
the initial gravitational instability of a galaxy after its formation or any
major perturbation. An unstable galaxy forms stars rapidly in an early time,
so its stellar populations will appear older than those in a more stable
galaxy. More massive galaxies are less stable than small galaxies with the
same gas fraction.  The different star formation histories in such galaxies
may account for the downsizing effect that star formation first occurs in big
galaxies at high redshift, while modern starburst galaxies are smaller
(\citealt{cowie96, poggianti04, ferreras04}), and thus more stable.

\subsection{Radial Distribution of Gas and Star Formation}

\begin{figure*}
\begin{center}
\includegraphics{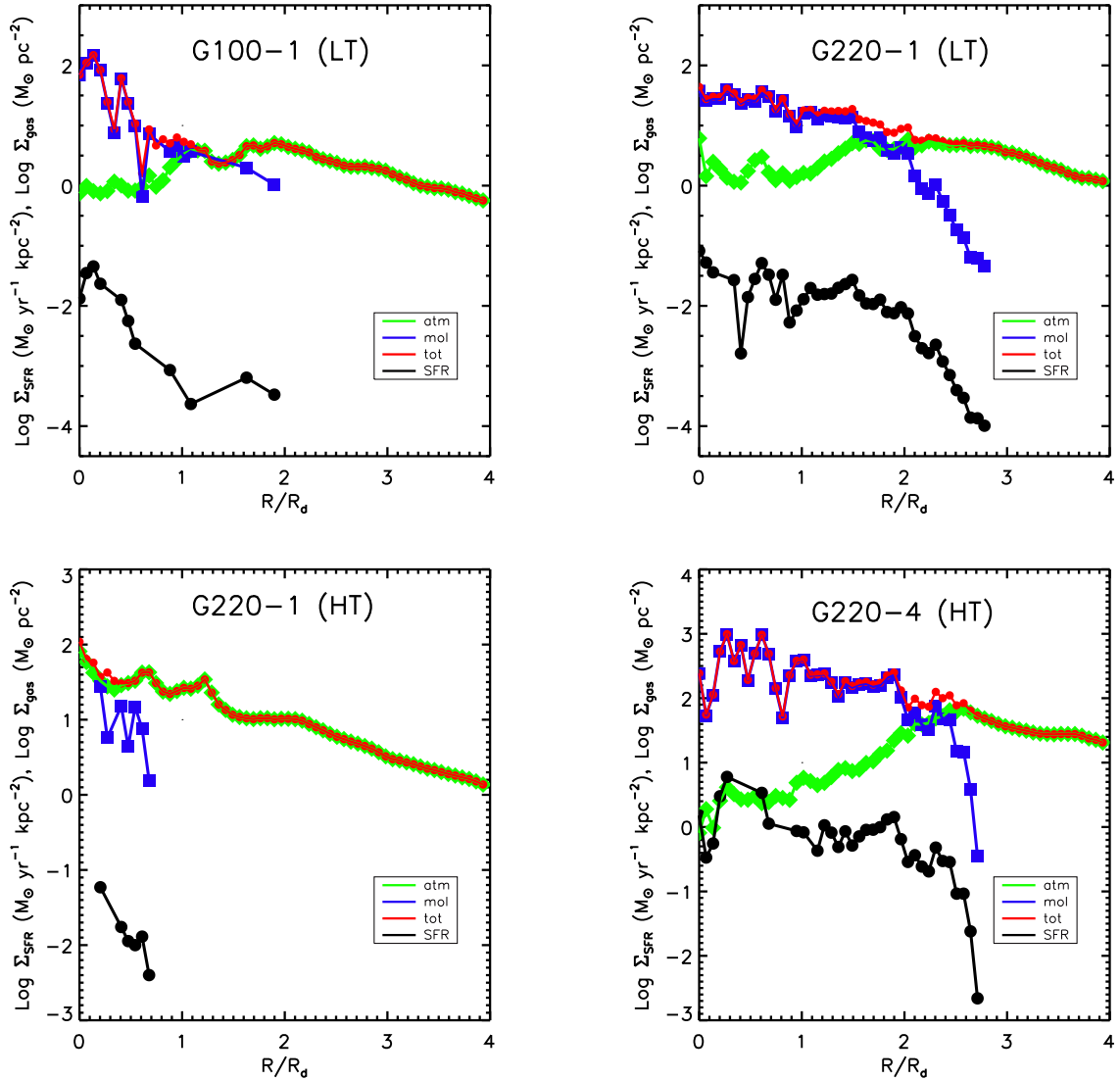}
\caption[Radial Profiles of SFR and Gas]{\label{fig_dist} Radial profiles of
atomic (SPH particles; {\em green dimond}), molecular (70\% of sink
particle mass; {\em blue square}), and total ({\em red dot}) gas surface density,
as well as SFR surface density ({\em black dot}). $R_{\rm d}$ is the radial 
disk scale length as given in Table~1. Note the solid lines are used to
connect the symbols.}   
\end{center}
\end{figure*}

Figure~\ref{fig_dist} shows the radial distribution of different gas
components and the SFR of selected models. The gas distribution and SFR are
calculated at the star formation timescale $\tau_{\rm SF}$ derived from the
fits given in Paper I. We assume that 70\% of the gravitationally collapsed,
high-density gas (as identified by sink particles) is in molecular
form. Similarly, we identify unaccreted gas as being in atomic form. The total
amount of gas is the sum of both components.

Figure~\ref{fig_dist} shows that the simulated disks have gas
distributions that are mostly atomic in the outer disk but dominated
by molecular gas in the central region. Observations by \citet{wong02}
and \citet{heyer04} show that the density difference between the
atomic and molecular components in the central region depends on the
size and gas fraction of the galaxy. For example, $\Sigma_{\rm H_2}$
in the center of NGC4321 is almost two orders of magnitude higher than
$\Sigma_{\rm H_I}$ \citep{wong02}, while in M33, the difference is
only about one order of magnitude \citep{heyer04}.  A similar relation
between the fraction of molecular gas and the gravitational
instability of the galaxy is seen in our simulations. In our most
unstable galaxies such as G220-4, the central molecular gas surface
density exceeds the atomic gas surface density by more than two orders
of magnitude, while in a more stable model like G220-1 (high~$T$),
the profiles of $\Sigma_{\rm H_2}$ and $\Sigma_{\rm H_I}$ are close to
each other within one disk scale length $R_{\rm d}$.

We also find a linear correlation between the molecular gas surface
density and the SFR surface density, as can be seen by their parallel
radial profiles in Figure~\ref{fig_dist}.  (In operational terms, we find
a correlation between the surface density in sink particles and the
rate at which they accrete mass.) \citet{gao04b} found a tight linear
correlation between the far infrared luminosity, a tracer of the star
formation rate, and HCN luminosity, in agreement with our result that star
formation rate and molecular gas have similar surface density profiles. 

The agreement between the simulations and observations supports our assumption
that both molecular gas and stars form by the gravitational collapse of high
density gas. Note, however, that we neglect recycling of gas from molecular
clouds back into the warm atomic and dissociated or ionized medium represented
by SPH particles in our simulation. Although it is possible that even that
reionized gas may still quickly collapse again if the entire region is
gravitationally unstable, this still constitutes an important limitation of
our models that will have to be addressed in future work. 

\section{GLOBAL SCHMIDT LAW}
\label{sec_global}
To derive the global Schmidt law from our models, we average $\Sigma_{\rm SFR}$ and
$\Sigma_{\rm gas}$ over the entire star forming region of each galaxy. We
define the star forming region following \citet{kennicutt89}, using a radius
chosen to encircle 80\% of the mass accumulated in sink particles (denoted $R_{80}$
hereafter). The SFR is taken from the SFR evolution curves at some chosen
time. As mentioned in \S~\ref{sec_dist}, 30\% of the mass of sink particles is
assumed to be stars, while the remaining 70\% of the sink particle mass
remains in molecular form. The atomic gas component is computed from the SPH
particles not participating in localized gravitational collapse, that
is, gas particles not accreted onto sink particles. The total is the
sum of atomic and molecular gas.

\begin{figure*}
\begin{center}
\includegraphics{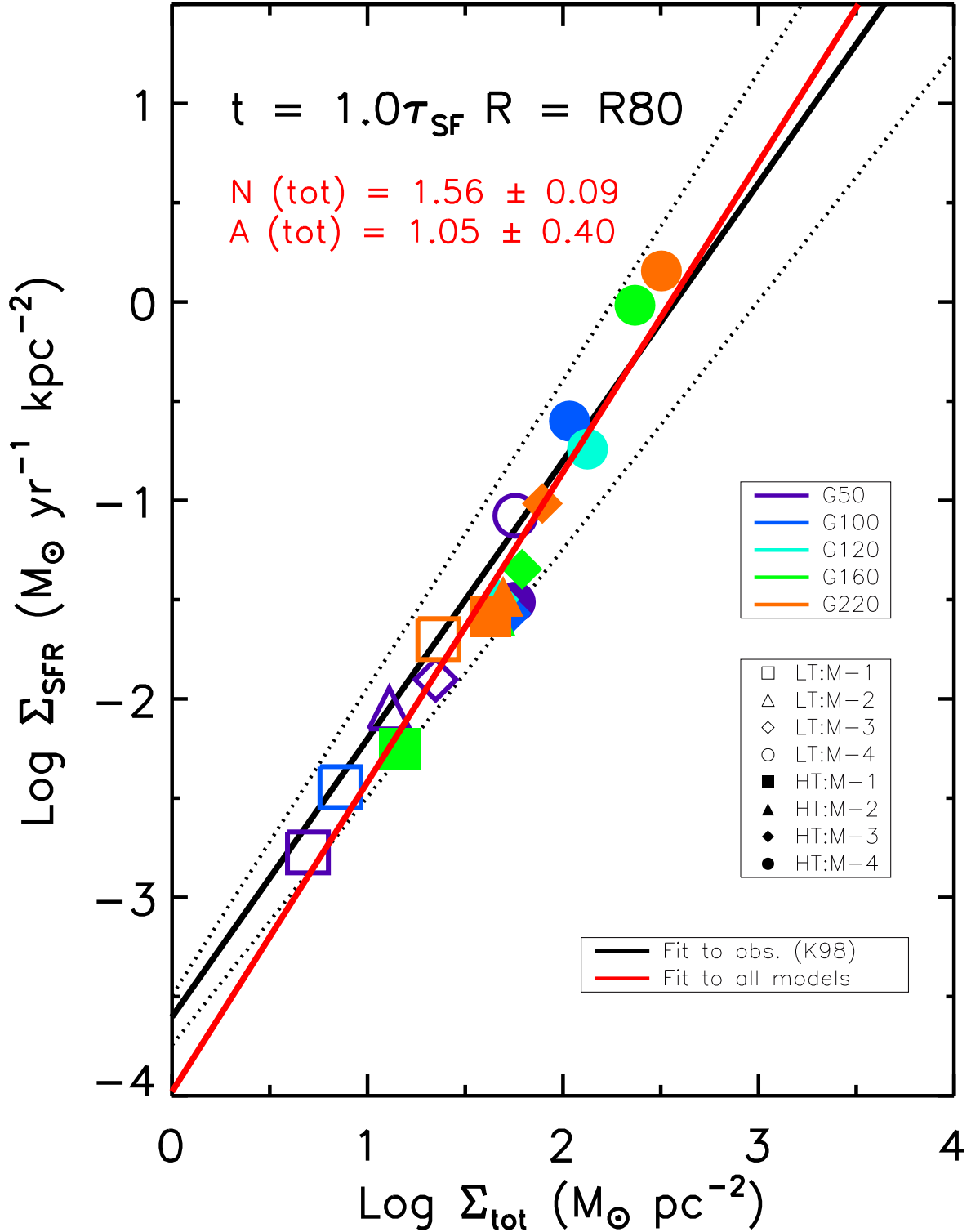}
\caption[Global Schmidt Law]{\label{fig_global} A comparison of the global
  Schmidt laws between our simulations and the observations. The red
  line is the least-square fit to the total gas of the simulated
  models, the black solid line is the best fit of observations from
  \citet{k98}, while the black dotted lines indicate the observational
  uncertainty. The color of the symbol indicates the rotational
  velocity for each model (see Table~1); labels from M-1 to M-4 are
  sub-models with increasing gas fraction; and open and filled symbols
  represent low and high~$T$ models, respectively.}
\end{center} 
\end{figure*}

Figure~\ref{fig_global} shows the global Schmidt laws derived from our
simulations at the star formation time $t=\tau_{\rm SF}$ as listed in
Table~1.  Note that for a few models this is just the maximum
simulated time, as indicated in Table~1. (Results from different times and
star formation regions are shown in the next section.) We fit the data to the total
gas surface density of the models listed in Table~1 (both low~$T$ and
high~$T$). A least-square fit to the models we have run gives a
simulated global Schmidt law
\begin{align} 
\label{gsl}
\Sigma_{\rm {SFR}} = (1.1 \pm 0.4 \times 10^{-4}  \mbox{ M}_{\odot} \mbox{
yr}^{-1} \mbox{ kpc}^{-2}) \notag \\
\times \left(\frac{\Sigma_{\rm {gas}}}{1\ \rm M_{\odot}\ {\rm pc}^{-2}}\right)^{1.56
  \pm 0.09} 
\end{align} 

For comparison, the best fit to the observations by \citet{k98} gives
a global Schmidt law for the total gas surface density in a sample
that includes both the normal and starburst galaxies of
\begin{align}
\label{eq_k98}
\Sigma_{\rm {SFR}} =  (2.5 \pm 0.7 \times 10^{-4} \mbox{ M}_{\odot} \mbox{
yr}^{-1} \mbox{ kpc}^{-2}) \notag \\
\times \left(\frac{\Sigma_{\rm {gas}}}{1 \ \rm 
M_{\odot}\ {\rm pc}^{-2}}\right)^{1.4 \pm 0.15}.
\end{align}

The global Schmidt law derived from the simulations agrees with the observed
slope within the observational errors, but has a normalization a bit lower
than the observed range.  There are three potential explanations for this
discrepancy. First, we have not weighted the fit by the actual distribution of
galaxies in mass and gas fraction.  Second, as we discuss in the next
subsection, we have not used models at different times in their lives weighted
by the distribution of lifetimes currently observed. Third, we have only
simulated isolated, normal galaxies. Our models therefore do not populate the
highest $\Sigma_{\rm SFR}$ values observed in \citet{k98}, which are all
starbursts occurring in interacting galaxies. These produce highly unstable
disks that undergo vigorous starbursts with high SFR \citep*[e.g.,][]{li04}.  
Our result is supported by \citet{boissier03}, who found a much deeper slope,
$N \sim 2$ in a sample of normal galaxies comparable to our more stable models.  
In the models, the local SFE $\epsilon_{\ell}$ is fixed at 30\%, independent
of the galaxy model. A change of the assumed value of $\epsilon_{\ell}$
changes the normalization but not the slope of our relation. For example, an
extremely high value of $\epsilon_{\ell}\sim 90$\% increases $A$ to $\sim 3.15 \times
10^{-4} \mbox{ M}_{\odot} \mbox{ yr}^{-1} \mbox{ kpc}^{-2}$, which is just
within the $1\sigma$ upper limit of the observation by \citet{k98}. If we
decrease $\epsilon_{\ell}$ to 10\%, then $A$ decreases to $\sim 0.7
\times 10^{-4} \mbox{ M}_{\odot} \mbox{ yr}^{-1} \mbox{ kpc}^{-2}$.
These fairly extreme assumptions still produce results lying within
the observed ranges \citep[e.g.,][]{wong02}, suggesting that our overall
results are insensitive to the exact value of the local SFE that we assume.

The SFR surface densities $\Sigma_{\rm SFR}$ change dramatically with
the gas fraction in the disk. The most gas-rich models (M-4, circles) have
the highest $\Sigma_{\rm SFR}$, while the models poorer in gas (M-1,
squares) have $\Sigma_{\rm SFR}$ two orders of magnitudes lower than
their gas-rich counterparts. Note that models with lower $\Sigma_{\rm
  SFR}$ tend to have slightly higher scatter, because in these models
fewer sink particles form, and they form over a longer period of time,
resulting in higher statistical fluctuations.

\begin{figure}
\begin{center}
\includegraphics[width=3.2in]{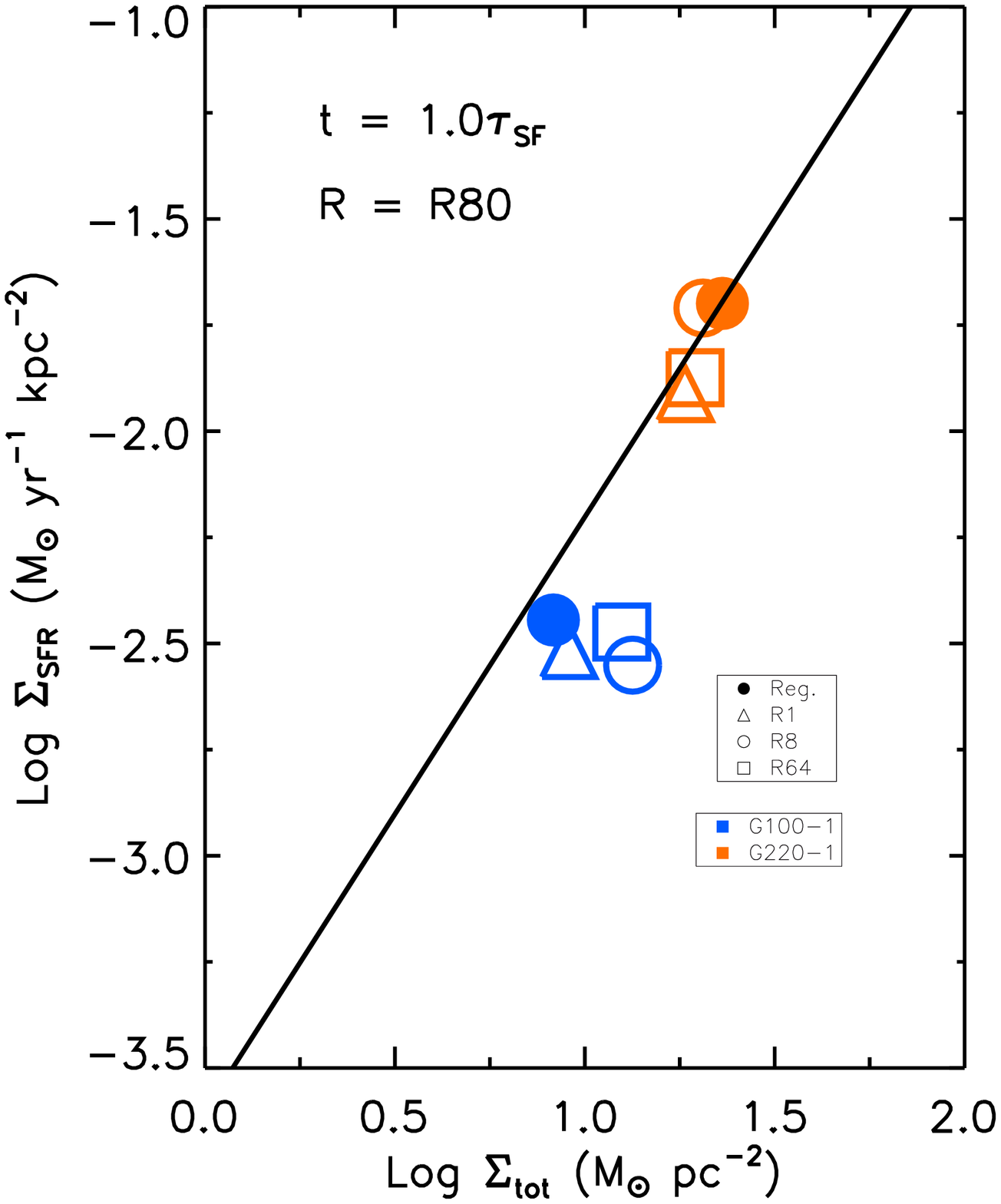}
\caption[Resolution Study of Global Schmidt Law]{\label{fig_res} Same as
  Figure~\ref{fig_global} but for the resolution study of low~$T$
  models of G100-1 ({\em blue}) and G220-1 ({\em orange}). Models with
  total particle number of $N_{\rm tot} = 10^5$ (R1; {\em open
    triangle}), $8\times 10^5$ (R8; {\em open circle}) and $6.4\times
  10^6$ (R64; {\em open square}) are shown. Models with regular
  resolution $N_{\rm tot} = 10^6$ (R10 {\em filled circle}) are also shown
  for comparison.}
\end{center} 
\end{figure}

A resolution study is shown in Figure \ref{fig_res} that compares the
global Schmidt law computed with different numerical resolutions. Runs
with different resolution converge within 10\% in both the
$\Sigma_{\rm SFR}$ and $\Sigma_{\rm gas}$. Although numerical
resolution affects the total mass collapsed, and the number and location of
fragments, as shown in Paper I, the SFR at $\tau_{\rm SF}$ seems to
be less sensitive to the numerical resolution.

\subsection{A Parameter Study}
\label{subsec_para}

\begin{figure}
\begin{center}
\includegraphics{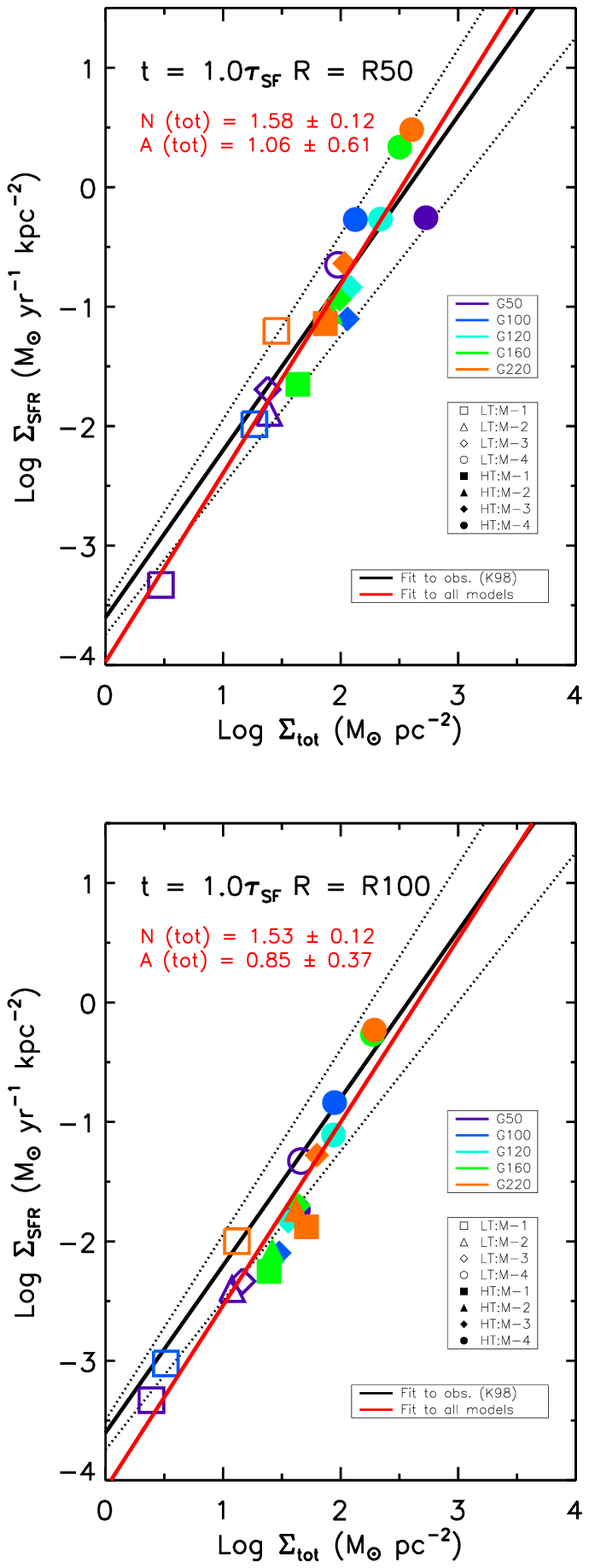}
\caption[Parameter Study of Global Schmidt Law on R]{\label{fig_par_r} Same as
  Figure~\ref{fig_global} but with different radii assumed for the
  star-forming region: (\textit{a}) $R = R_{50}$, and (\textit{b}) $R
  = R_{100}$.  The time is fixed at $t =\tau_{\rm SF}$.  }
\end{center} 
\end{figure}

In order to test how sensitive the global Schmidt law is to the radius
$R$ and the time $t$ chosen to measure it, we carry out a parameter
study changing both $R$ and $t$ individually. To maintain consistency
with the previous section, we continue to assume a constant local SFE
$\epsilon_{\ell}=30$\%.  Figure~\ref{fig_par_r} compares the global
Schmidt laws in total gas at different radii for the star-forming
region $R = R_{50}$ and $R_{100}$ (encircling 50\% and 100\% of the
newly formed star clusters), while the time is fixed to $t= \tau_{\rm
  SF}$. We can see that the case with $R_{50}$ has larger scatter than
that with $R_{100}$. This is due to the larger statistical
fluctuations caused by the smaller number of star clusters within this
radius. The global Schmidt law with $R = R_{100}$ is almost identical
to that with $R = R_{80}$ shown in Figure~\ref{fig_global}.

\begin{figure}
\begin{center}
\includegraphics{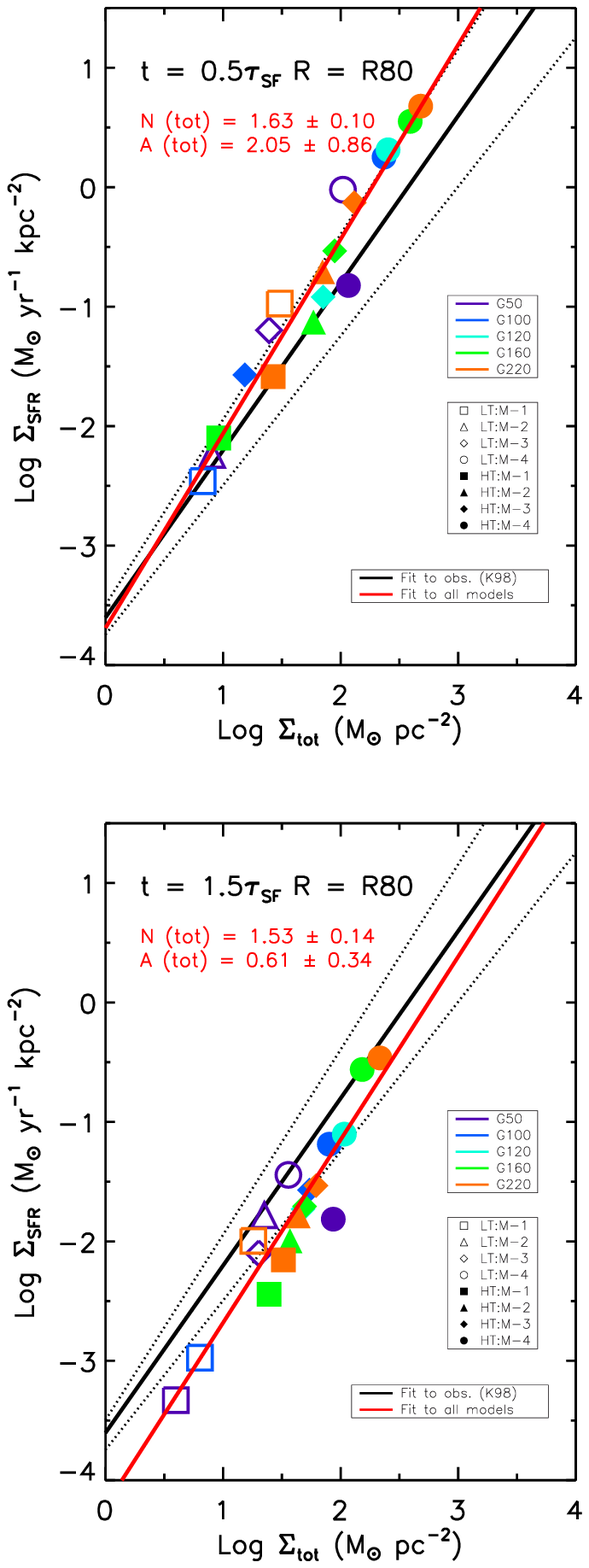}
\caption[Parameter Study of Global Schmidt Law on t]{\label{fig_par_t} Same as
  Figure~\ref{fig_global} but derived at different times (\textit{a})
  $t = 0.5\ \tau_{\rm SF}$ and (\textit{b}) $t = 1.5\ \tau_{\rm SF}$.
  The radius for the star formation region is fixed at $R_{80}$.}
\end{center} 
\end{figure}

Figure \ref{fig_par_t} compares the global Schmidt laws in total gas
at different times $t= 0.5 \tau_{\rm SF}$ and $t=1.5 \tau_{\rm SF}$
with the star formation radius fixed to $R_{80}$. Compared to the
$t=\tau_{\rm SF}$ case, models in the $t= 0.5 \tau_{\rm SF}$ case 
have higher $\Sigma_{\rm SFR}$, because the SFR drops almost exponentially
with time (\S~\ref{subsec_sfr}). Similarly, models in the $t= 1.5 \tau_{\rm
SF}$ case shift to the lower right. Nevertheless, data derived from different
times appear to preserve the power-law index of the Schmidt law, just
differing in the normalizations.  

The global Schmidt law presented by \citet{li05b} was derived at a time when
the total mass of the star clusters reached 70\% of the maximum collapsed
mass, which is close to 1.0 $\tau_{\rm SF}$ in many models.  The time interval
$\Delta t$ used to calculate the SFR was the time taken to grow from 30\%
to 70\% of the maximum collapsed mass, rather than the $\Delta t=50$~Myr used
here. Nevertheless, the results presented here also agree well with those in
\citet{li05b}. 

Our parameter study demonstrates that the global Schmidt law depends
only weakly on the details of how it is measured. The small scatter
seen in \citet{k98} does suggest that additional physics not included
in our modeling may be important.  We should keep in mind that since
we do not treat gas recycling, our models are valid only within one
gas consumption time $\tau_{\rm SF}$. The evolution after
that may become unrealistic as most of the gas is locked up in the
sinks. Nevertheless, our results suggest that the Schmidt law is a
universal description of gravitational collapse in galactic disks.

\subsection{Alternative Global Star Formation Laws}
\label{subsec_alt}

The existence of a well-defined global Schmidt law suggests that the
star formation rate depends primarily on the gas surface density. As
shown by several authors (e.g., \citealt{quirk72, larson88, kennicutt89,
  elmegreen94a, k98}), a simple picture of gravitational collapse on a
free-fall timescale $\tau_{\rm ff} \propto \rho^{-1/2}$ qualitatively
produces the Schmidt law.  Assuming the gas surface density is
directly proportional to the midplane density, $\Sigma_{\rm gas}
\propto \rho$, it follows that $\Sigma_{\rm SFR} \propto \Sigma_{\rm
  gas}/\tau_{\rm ff} \propto \Sigma_{\rm gas}^{3/2}$. This suggests
that the Schmidt law reflects the global growth rate of gas density
under gravitational perturbations.

An alternative scenario that uses the local dynamical timescale has been
suggested by several groups (e.g. \citealt{wyse86, wyse87, silk97,
elmegreen97, hunter98, tan00}). In particular, \citet{elmegreen97} and
\citet{hunter98} proposed a kinematic law that accounts for the stabilizing
effect of rotational shear, in which the global SFR scales with the angular
velocity of the disk, 
\begin{equation}
\label{eq_torb}
\Sigma_{\rm SFR} \propto \frac{\Sigma_{\rm gas}}{t_{\rm orb}} \propto
\Sigma_{\rm gas}\Omega 
\end{equation}
where $t_{\rm orb}$ is the local orbital timescale and $\Omega$ is the orbital
frequency. \citet{k98} gave a simple form for the kinematical law,  
\begin{equation}
\label{eq_omega}
\Sigma_{\rm SFR} \simeq 0.017\ \Sigma_{\rm gas}\Omega  
\end{equation}
with the normalization corresponding to a SFR of 21\% of the gas mass per orbit at the
outer edge of the disk. 

\begin{figure*}
\begin{center}
\includegraphics{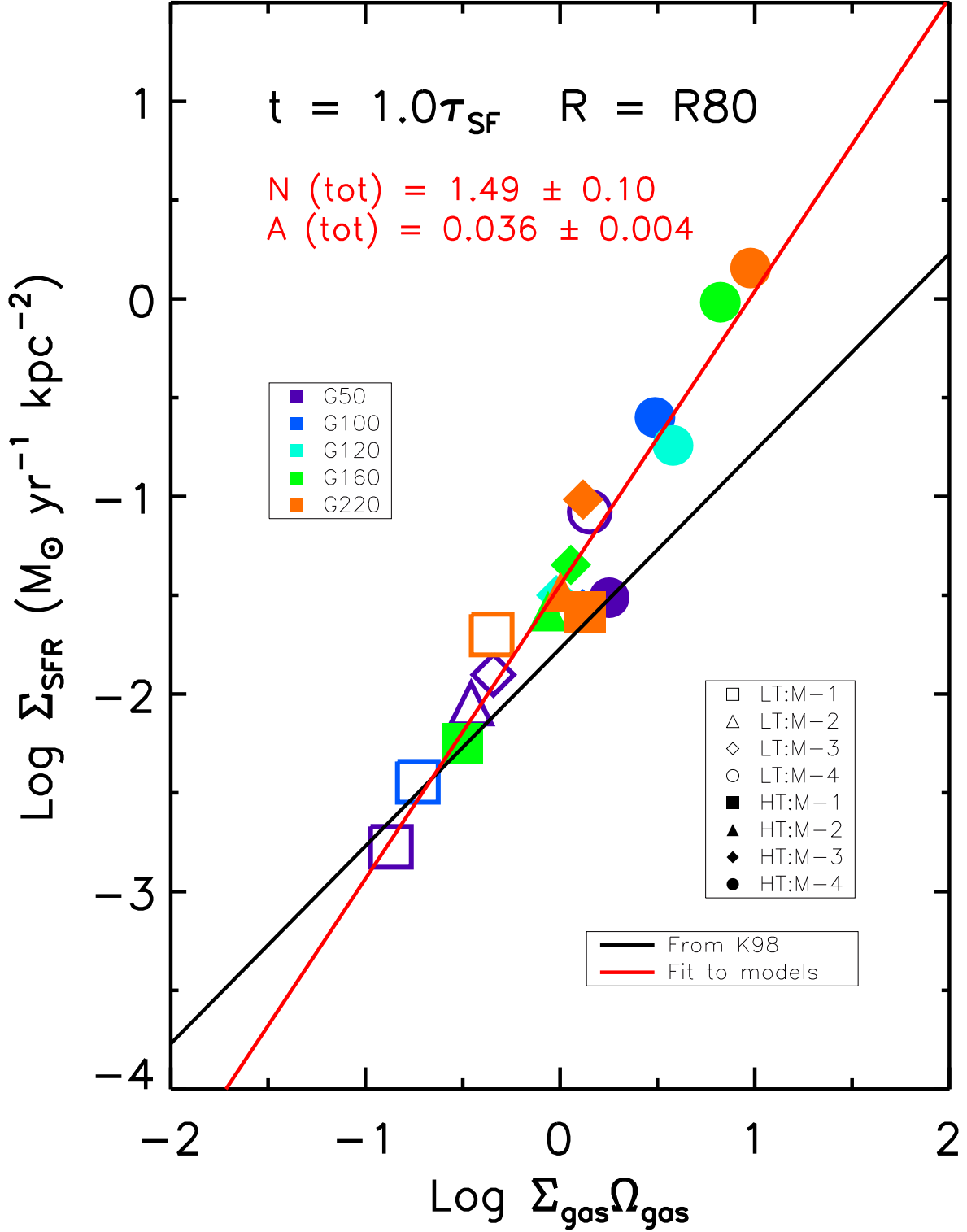}
\caption[Correlation between $\Sigma_{\rm gas}$ and $\Omega_{\rm
gas}$]{\label{fig_tdyn} Relation between $\Sigma_{\rm SFR}$ and $\Sigma_{\rm
    gas} \Omega$. The legends are the same as in Figure~\ref{fig_global}. The
black solid line is the linear relationg derived from the observations by 
\citet{k98}, as given in our equation (\ref{eq_omega}).}
\end{center} 
\end{figure*}

For our analysis, we follow \citet{k98} and define $t_{\rm orb} = 2\pi
R/V(R) = 2\pi /\Omega(R)$, where $V(R)$ is the rotational velocity at
radius $R$. We use the initial rotational velocity, which should not
change much with time as it depends largely on the potential of the
dark matter halo. Figure~\ref{fig_tdyn} shows the relationship between
$\Sigma_{\rm SFR}$ and $\Sigma_{\rm gas}\Omega$ in our models. The
densities of SFR $\Sigma_{\rm SFR}$ and total gas $\Sigma_{\rm gas}$
are calculated the same way as in Figure~\ref{fig_global} at 1.0
$\tau_{\rm SF}$ and $R_{80}$, and $\Omega(R)$ is calculated by using
the initial total rotational velocity at $R_{80}$. A least-square fit to
the data gives $\Sigma_{\rm SFR} = (0.036 \pm 0.004)\times
\left(\Sigma_{\rm gas}\Omega\right)^{1.49 \pm 0.1}$. This correlation
has a steeper slope than the linear law given in
equation~(\ref{eq_omega}), suggesting a discrepancy between the
behavior of our models and the observed kinematical law.

However, \citet{boissier03} recently reported a slope of $\sim$ 1.5
for the kinematical law from observations of 16 normal disk galaxies,
in agreement with our results.  Examination of Figure~7 in \citet{k98}
also shows that the normal galaxies, considered alone, seem to have a
steeper slope than the galactic nuclei and starburst galaxies.
\citet{boissier03} suggested several reasons for the discrepancy, the
most important one being the difference between their sample of normal
galaxies and the sample of \citet{k98} including many starburst
galaxies and galaxy nuclei.  More simulations, and models with higher
SFR such as galaxy mergers are necessary to test this hypothesis.

\subsection{A New Parameterization}
\label{subsec_new}

The global Schmidt law describes global collapse in the gas disk. It
does not seem to depend on the local star formation process. From
Paper I and \citet{li05b} we know that $\Sigma_{\rm SFR}$
correlates tightly with the strength of gravitational instability
(see also \citealt{mk04, khm00, hmk01}). Here we quantify this
correlation, using the Toomre Q parameter to measure the strength of
instability.

\begin{figure}
\begin{center}
\includegraphics{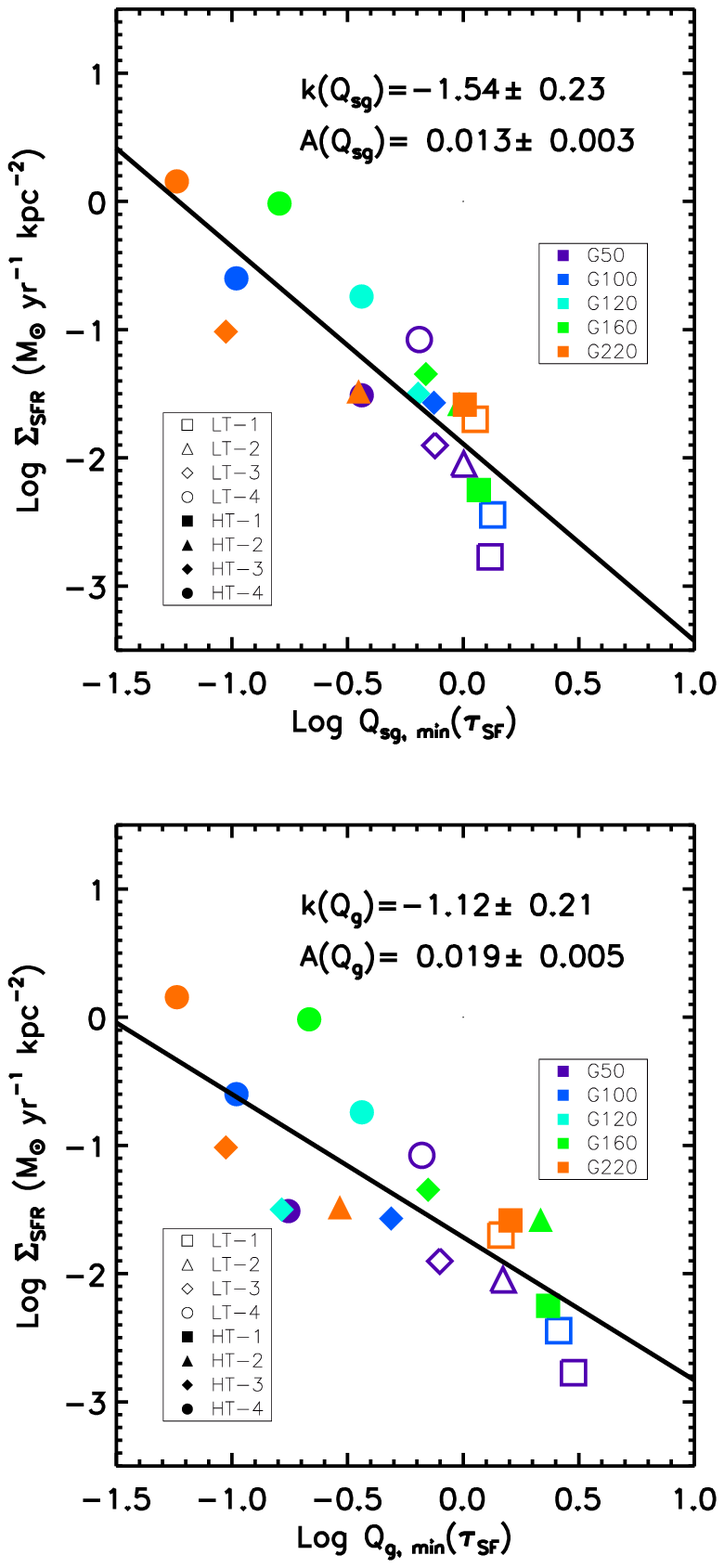}
\caption[Correlation between $\Sigma_{\rm SFR}$ and $Q_{sg,{\rm
  min}}(t)$]{\label{fig_sfr_q} Correlations between $\Sigma_{\rm SFR}$
  and the minimum value of ({\em a}) the Toomre parameter for stars
  and gas $Q_{sg}(t)$ and ({\em b}) the Toomre parameter for gas
  $Q_g(t)$ at $t = 1\ \tau_{\rm SF}$. The legends are the same
  as in Figure~\ref{fig_global}. The solid black lines are the
  least-square fits to the data given in equation~(\ref{eq_sfr_q_fit}).}
\end{center} 
\end{figure}

Figure~\ref{fig_sfr_q} shows the correlation between $\Sigma_{\rm
  SFR}$ and the local gravitational instability parameters $Q_{\rm
  min}$. The parameters $Q_{\rm min}$ are minimum values of the Toomre
$Q$ parameters for gas $Q_{g,{\rm min}}(t)$, and the combination of
stars and gas $Q_{sg,{\rm min}}(t)$ at a given time $t$, respectively.
To obtain the $Q$ parameters, we follow the approach of
\citet{rafikov01}, as described in equations (1)--(3) of Paper I. We
divide the entire galaxy disk at time $t$ into 40 annuli, calculate
the $Q$ parameters in each annulus, then take the minimum. In the
plots, the time when $\Sigma_{\rm SFR}$ is computed is $t = \tau_{\rm
  SF}$. This correlation does not change significantly with time, but the
  scatter becomes larger at later times because the disk becomes more clumpy,
  which makes the calculation of $Q_{g,{\rm min}}(t)$ more difficult (see
  below).

There is substantial scatter in the plots, at least partly caused 
by the clumpy distribution of the gas.  Equations (1)--(3) for the $Q$
parameters in Paper I are derived for uniformly distributed gas, such
as in our initial conditions. As the galaxies evolve, the gas forms
filaments or spiral arms probably leading to the fluctuations seen.  
The least-square fits to the data shown in Figure~\ref{fig_sfr_q}
give
\begin{align}
\label{eq_sfr_q_fit}
\Sigma_{\rm SFR} = \left(0.013 \pm 0.003\  \mbox{ M}_{\odot} \mbox{
yr}^{-1} \mbox{ kpc}^{-2}\right) \notag \\
\times \left[Q_{sg,{\rm min}}(\tau_{\rm SF})\right]^{-1.54 \pm 0.23} 
\end{align}

\begin{align}
\Sigma_{\rm SFR} = \left(0.019 \pm 0.005\  \mbox{ M}_{\odot} \mbox{
yr}^{-1} \mbox{ kpc}^{-2} \right) \notag \\
\times \left[Q_{g,{\rm min}}(\tau_{\rm SF})\right]^{-1.12 \pm 0.21}  
\end{align}

If we take a first-order approximation, $Q_{sg} \propto 1\ /\Sigma_{\rm gas}$,
then equation (\ref{eq_sfr_q_fit}) gives $\Sigma_{\rm SFR} \propto \Sigma_{\rm
gas}^{1.54}$ at $t=\tau_{\rm SF}$, agreeing very well with the observations. 
The slopes derived from $Q_g$ appear to be lower than those derived from
$Q_{sg}$, but are still within the slope range observed. 

Keep in mind that the local instability is a non-linear interaction
between the stars and gas, and so is much more complicated than the
linear stability analysis presented here. Also, the instability of the
entire disk at a certain time is not fully represented by the minimum values
of the Q parameters we employ here, although they do represent the region of
fastest star formation. These factors limit our ability to derive the global
Schmidt law directly from the instability analysis.

\section{LOCAL SCHMIDT LAW}
\label{sec_local}
The relationship between surface density of SFR $\Sigma_{\rm SFR}$ and gas
density $\Sigma_{\rm gas}$ can also be measured as a function of radius 
within a galaxy, giving a local Schmidt law. Observations by \citet{wong02}
and \citet{heyer04} show significant variations in both the indices $N$ and
normalizations $A$ of the local Schmidt laws of individual galaxies. For
example, \citet{heyer04} show that M33 follows the law 

\begin{align}
\Sigma_{\rm {SFR}} =  (0.0035 \pm 0.066\  \mbox{
  M}_{\odot} \mbox{ yr}^{-1} \mbox{ kpc}^{-2}) \notag \\
\times \left({\Sigma_{\rm {tot}}}/{1\ 
\rm M_{\odot}\ {\rm pc}^{-2}}\right)^{3.3 \pm 0.07},
\end{align} 
while \citet{wong02} show that $N$ has a range of 1.23--2.06 in their sample. 

To derive local Schmidt laws, we again divide each individual galaxy into
40 radial annuli within 4 $R_{\rm d}$, then compute $\Sigma_{\rm gas}$ and
$\Sigma_{\rm SFR}$ in each annulus. The SFR is measured at $t = \tau_{\rm SF}$
as in \S~\ref{sec_global}.  For models where $\tau_{\rm SF} > 3$~Gyr or is
beyond the simulation duration, the maximum simulated timestep is used
instead, as listed in Table~1.

\subsection{Star Formation Thresholds}
\label{subsec_threshold}

\begin{figure*}
\begin{center}
\includegraphics{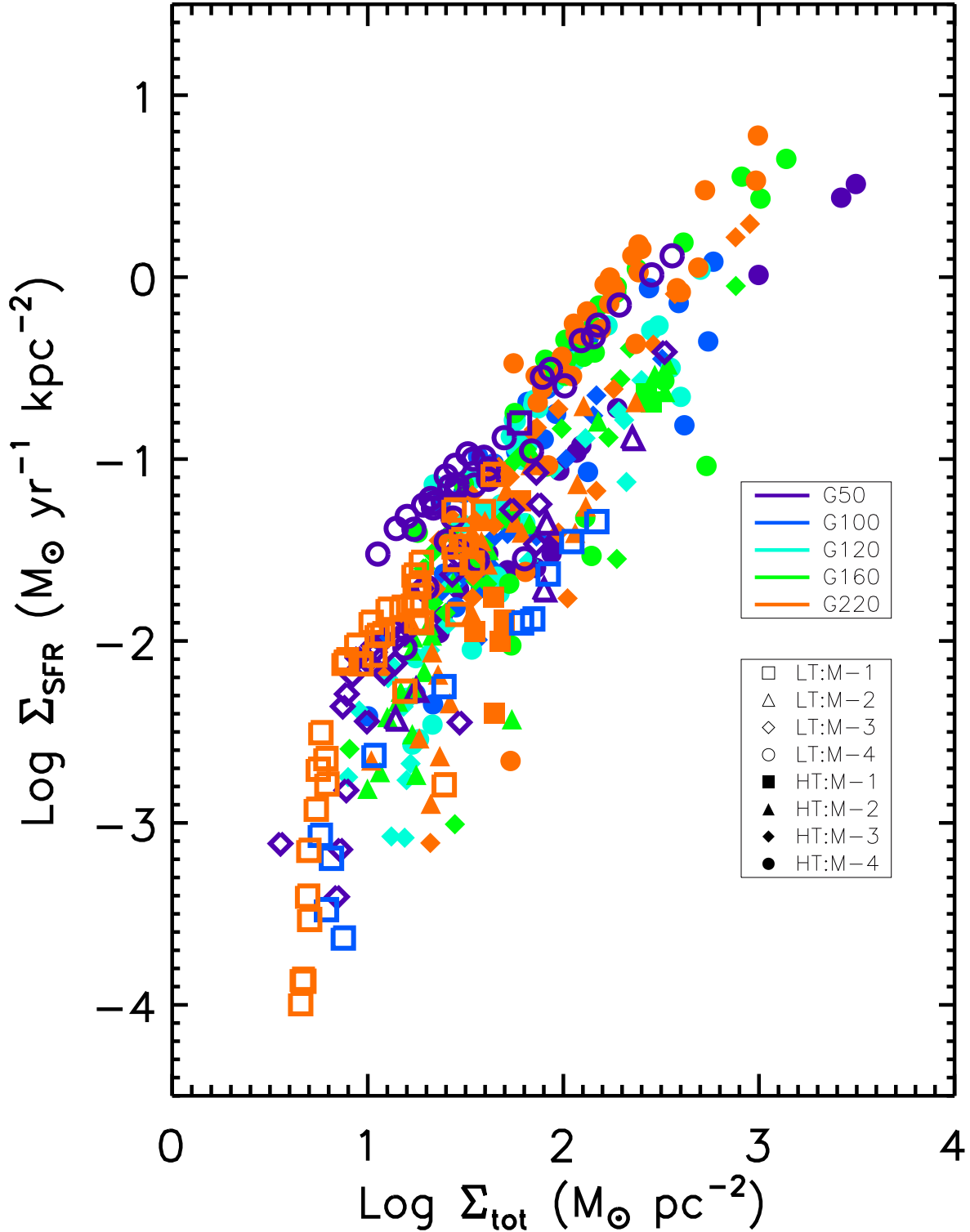}
\caption[Local Schmidt Laws of All Models]{\label{fig_local}
  Local Schmidt laws of all models with $\tau_{\rm SF} < 3$~Gyr. The
  legends are the same as in Figure~\ref{fig_global}: the color of
  the symbol indicates the rotational velocity for each model as given
  in Table~1, the shape indicates the sub-model classified by gas
  fraction, and open and filled symbols represent low and high $T$
  models, respectively.}
\end{center} 
\end{figure*}

Figure~\ref{fig_local} shows the relation between $\Sigma_{\rm SFR}$
and $\Sigma_{\rm gas}$ correlations for all models in the simulations
that form stars in the first 3 Gyr. With such a large number of models
in one plot, it is straightforward to characterize the general
features, and to compare with the observations shown in Figure~3 of
\citet{k98}.  Similar to the individual galaxies in \citet{k98} and
\citet{martin01}, each model here shows a tight $\Sigma_{\rm
  SFR}$--$\Sigma_{\rm gas}$ correlation, the local Schmidt law.
However, $\Sigma_{\rm SFR}$ drops dramatically at some gas surface
density.  This is a clear indication of a star-formation threshold.

We therefore define a threshold radius $R_{\rm th}$ as the radius that
encircles 95\% of the newly formed stars.  The gas surface density at
the threshold radius $R_{\rm th}$ in Figure~\ref{fig_local} has a
range from $\sim 4 \rm {M}_{\odot}\ {pc}^{-2}$ for the relatively stable model
G220-1 (low-$T$) to $\sim 60 \rm {M}_{\odot}\ {pc}^{-2}$ for the most unstable
model G220-4 (high-$T$). Note that in some galaxies, there are also smaller
dips in SFR at higher density. \citet{martin01} suggest that rotational
shearing can cause an inner star-formation threshold. However, the
inner dips in our simulations are likely due to the lack of accretion
onto sink particles in the simulations after most of the gas in the
central region has been consumed. Further central star formation in real
galaxies would occur due to gas recycling, which we neglect, and, probably
more important, after interactions with other galaxies.

\begin{figure}
\begin{center}
\includegraphics[width=3.2in]{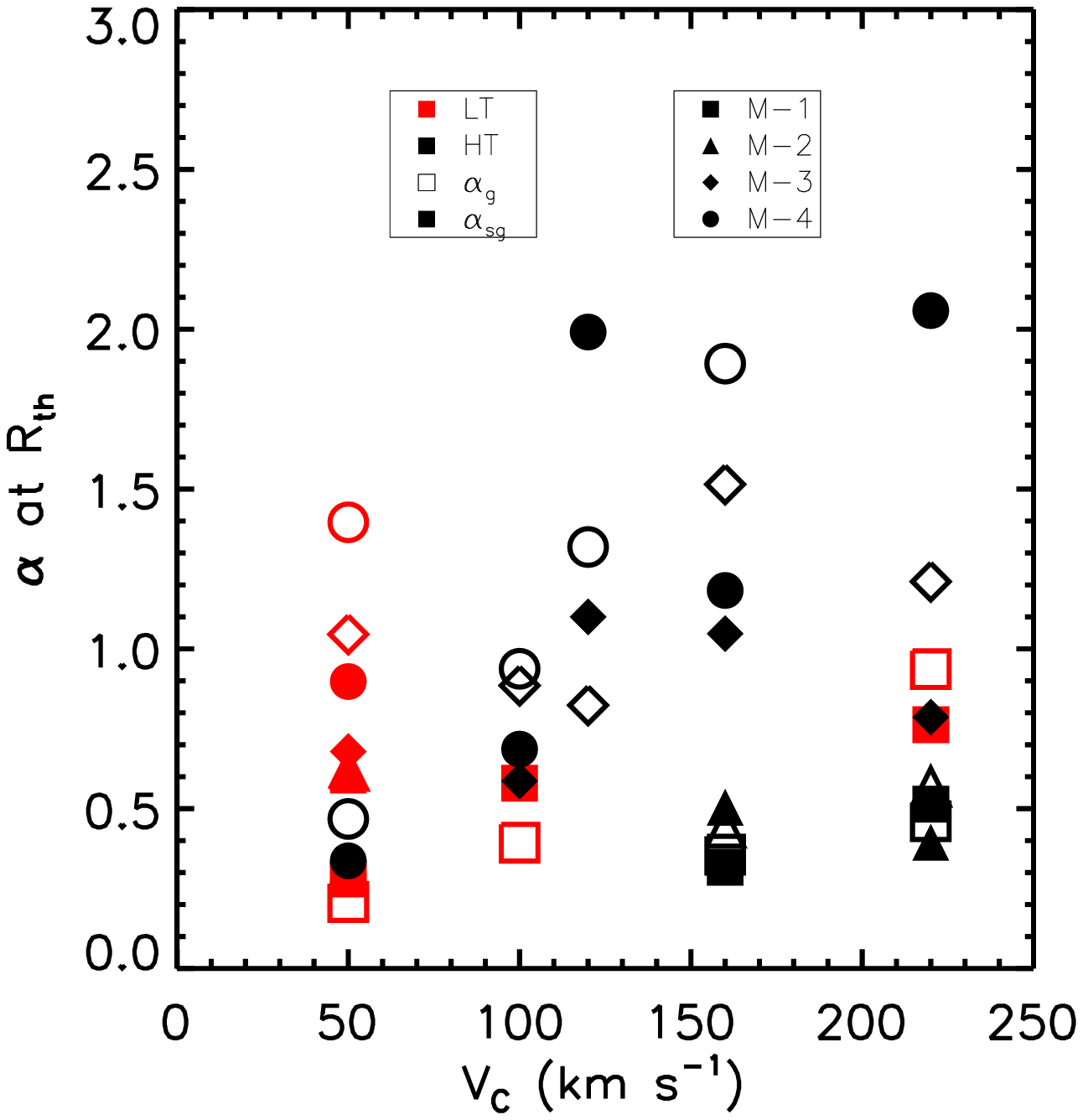}
\caption{\label{fig_local_alpha} The dimensionless parameter giving ratio
  of star formation threshold surface density to critical surface
  density for Toomre instability $\alpha_Q = 1/Q_{\rm th}$ is shown
  for low-$T$ (in red) and high-$T$ (in black) models. Open symbols
  give values derived using gas only, $\alpha_g = 1/Q_{g,{\rm th}}$,
  while filled symbols give values derived using both stars and gas,
  $\alpha_{sg} = 1/Q_{sg,{\rm th}}$. Gas fraction of disks increases
  from M-1 to M-4 (see Table~1).}
\end{center} 
\end{figure}

In the analysis of observations, a dimensionless parameter, $\alpha_Q =
\Sigma_{\rm th}/\Sigma_{\rm crit} = 1/Q$ has been introduced to relate the
star formation threshold to the Toomre unstable radius
\citep{kennicutt89}. The critical radius is usually defined as the 
radius where $Q_{g} = 1$. With a sample of 15 spiral galaxies,
\citet{kennicutt89} found $\alpha_Q \simeq 0.63$ by assuming a
constant effective sound speed (the velocity dispersion) of the gas
$c_s = 6$~km~s$^{-1}$. This result was confirmed by \citet{martin01} with a
larger sample of 32 well-studied nearby spiral galaxies, who reported a range
of $\alpha_Q \sim 0.3$--1.2, with a median value of 0.69.  However,
\citet{hunter98} found $\alpha_Q \simeq 0.25$ for a sample of irregular
galaxies with $c_s = 9$~km~s$^{-1}$. As pointed out by \citet{schaye04}, this
derivation of $\alpha_Q$ depends on the assumption of $c_s$.
The values of $\alpha_Q$ derived from our models using their actual values of
$c_s$ as shown in Figure~\ref{fig_local_alpha}. We find that the value
of $\alpha_Q$ depends not only on the gas sound speed, but also on the
gas fraction of the galaxy. For models with the same rotational velocity and
gas fraction, lower gas sound speed results in a higher value of
$\alpha_Q$. For models with the same total mass and sound speed, higher gas
fraction leads to higher $\alpha_Q$. The gas-poor models in our simulations
($f_{\rm g}=20\%$) have a range of $\alpha_Q \sim$ 0.2--1.0, agreeing roughly
with observations. This again may reflect the relative stability of the nearby
galaxies in the observed samples.

There are several theoretical approaches to explain the presence of
star formation thresholds. \citet{martin01} suggest that the
gravitational instability model explains the thresholds well,
with the deviation of $\alpha_Q$ from one simply due to the
non-uniform distribution of gas in real disk galaxies.
\citet{hunter98} proposed a shear criterion for star forming dwarf
irregular galaxies, as they appear to be sub-critical to the Toomre
criterion. \citet{schaye04} modeled the thermal and ionization
structure of a gaseous disk. He found the critical density is about
$\Sigma_{\rm crit} \sim3$--10~$\rm {M}_{\odot}\ \rm {pc}^{-2}$ with
a gas velocity dispersion of $\sim$~10~km~s$^{-1}$, and argued that
thermal instability determines the star formation threshold in the
outer disk. Our models suggest that the threshold depends on the
gravitational instability of the disk. The derived $\Sigma_{\rm crit}$
and $\alpha_Q$ from our stable models ($Q_{sg,{\rm min}} > 1$) agree
well with observations, supporting the arguments of \citet{martin01}.

\subsection{Local 
Correlations Between Gas and Star Formation Rate}
\label{subsec_local} 
\begin{figure*}
\begin{center}
\includegraphics{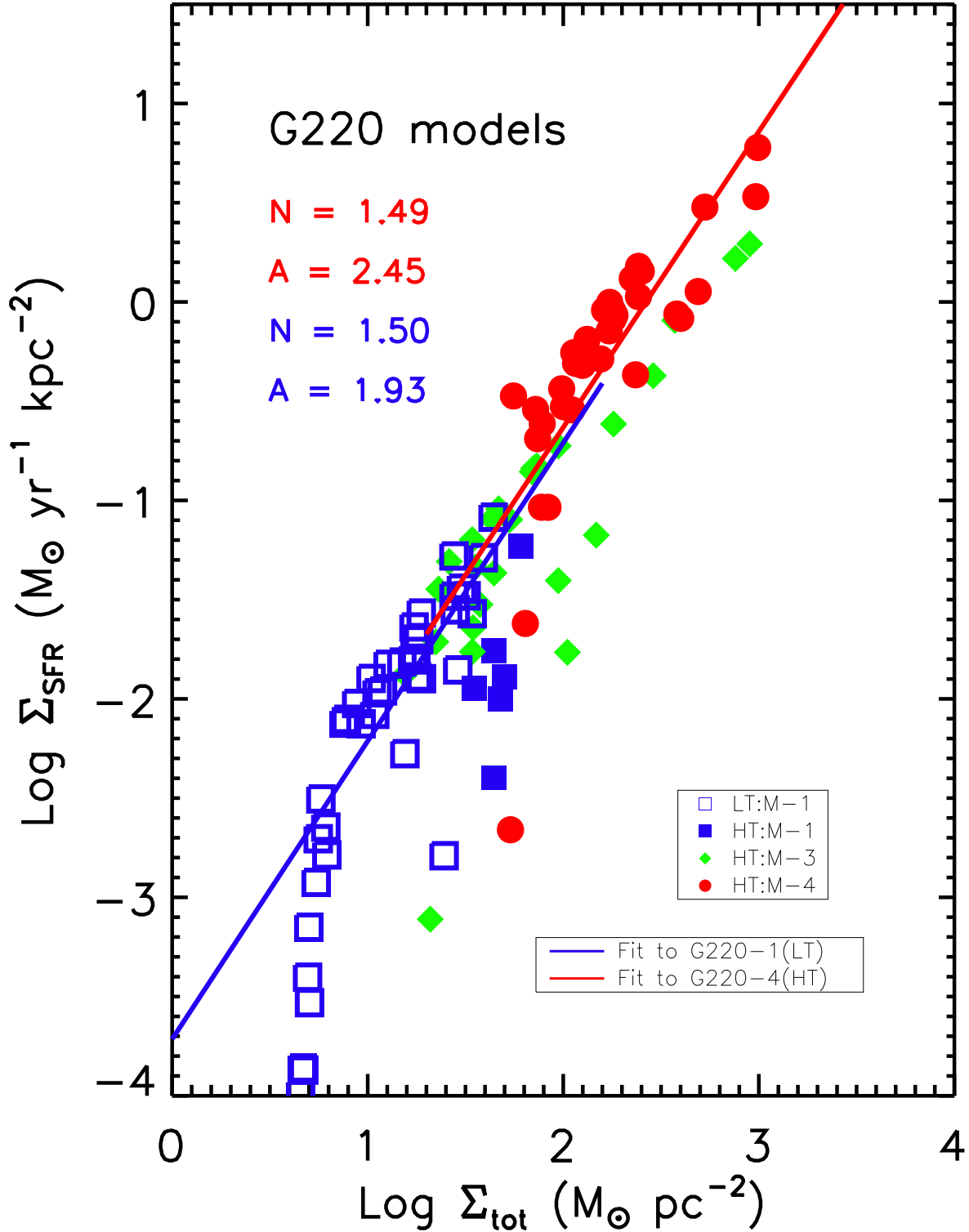}
\caption[Local Schmidt Laws of G220 models]{\label{fig_local_g220}
  Comparisons of local Schmidt laws of galaxy models with the same rotational
  velocity of 220 km s$^{-1}$ but different gas fractions and sound speeds, as
  indicated in the legend. The solid lines are least-square fits to data
  points within the threshold radius $R_{\rm th}$.}  
\end{center} 
\end{figure*}

We fit the local Schmidt law to the total gas surface density within $R_{\rm
  th}$, as demonstrated in Figure~\ref{fig_local_g220}. The models in
Figure~\ref{fig_local_g220} all have the same rotational velocity of
220~km~s$^{-1}$ but different gas fractions and sound speeds. The
local Schmidt laws of these models vary only slightly in slope and
  normalization.

\begin{figure}
\begin{center}
\includegraphics{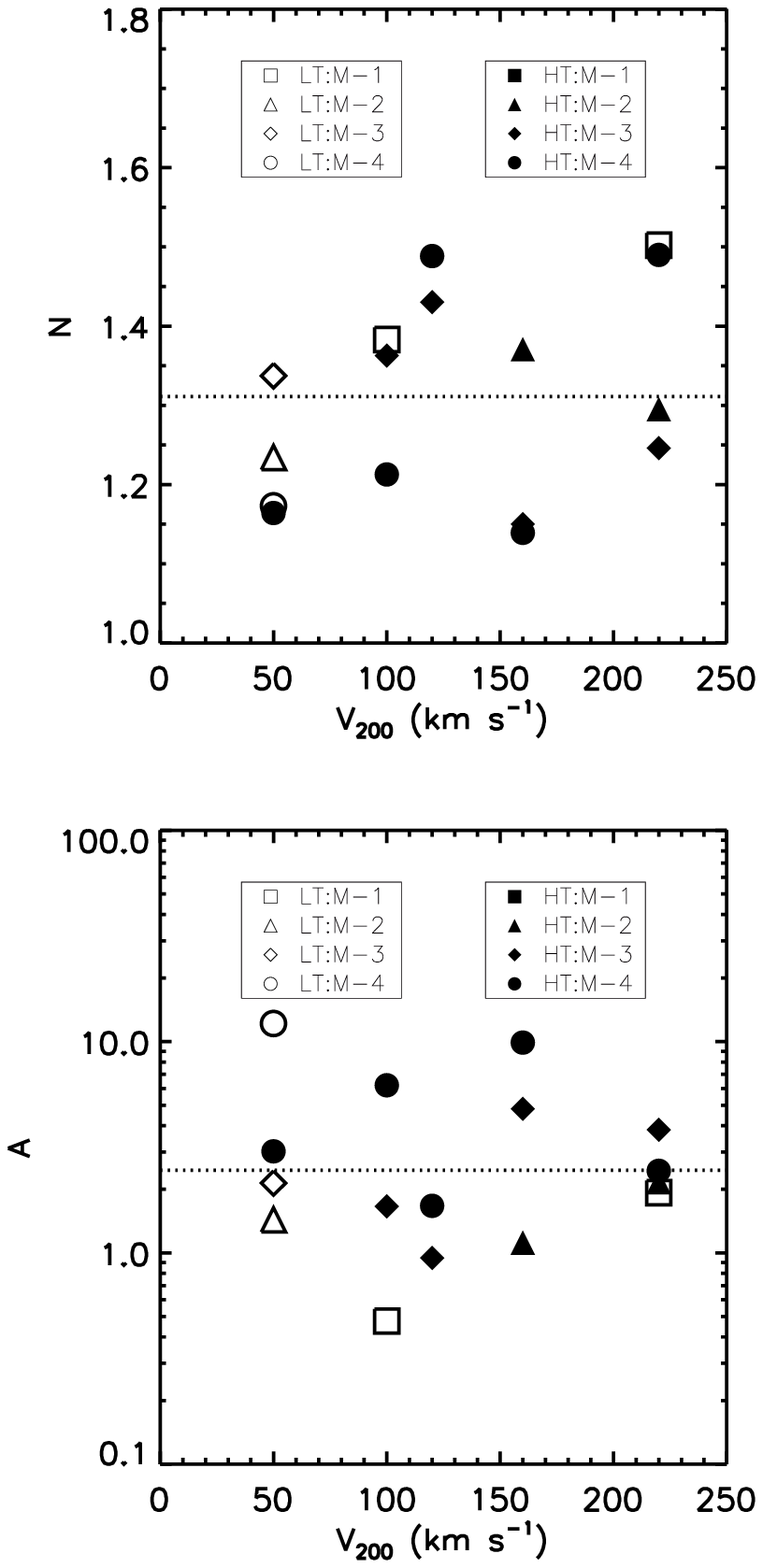}
\caption[Slopes and Normalizations of Local Schmidt
Laws]{\label{fig_local_fit} Local Schmidt law ({\em a}) slope $N$ and
  ({\em b}) normalization $A$ for all models with $\tau_{\rm SF} <
  3$~Gyr. The fit is to the total gas surface density $\Sigma_{\rm
    tot}$ within the threshold radius $R_{\rm th}$.  The symbols
  represent different models, as indicated in the legends.  The dotted
  line in each panel indicates the linearly averaged value across the
  models shown.}
\end{center} 
\end{figure}

Figures \ref{fig_local_fit}{\em a} and \ref{fig_local_fit}{\em b}
compare the slope $N$ and normalization $A$ of the local Schmidt laws for all
models in Table~1 that form stars in the first 3 Gyr. The slope of the fit to
the total gas in Figure~\ref{fig_local_fit}a varies from about 1.2 to
1.7. Larger galaxies tend to have larger $N$. However, the average slope is
around 1.3, agreeing reasonably well with that of the global Schmidt law.

There is substantial fluctuation in the normalization of fits to the
total gas, as shown in Figure~\ref{fig_local_fit}({\em b}). The
variation is more than an order of magnitude, with gas-rich models tending
to have high $A$.  However, the average value of $A$ settles around
2.2, agreeing surprisingly well with that of the global Schmidt law.
Overall, the averaged local Schmidt law gives:
\begin{align} 
\label{eq_local}
\Sigma_{\rm {SFR}} = (2.46 \pm 1.62 \times 10^{-4}\ \mbox{ M}_{\odot}\  \mbox{
yr}^{-1} \mbox{ kpc}^{-2}) \notag \\
\times \left(\frac{\Sigma_{\rm {tot}}}{1 \ \rm M_{\odot}\ {\rm
    pc}^{-2}}\right)^{1.31 \pm 0.15}  
\end{align}
 
The averaged local Schmidt law is very close to the global Schmidt law in
\S~\ref{sec_global}. 

The {\em average} slope in equation~(\ref{eq_local}) is rather smaller than
the value of $N = 3.3$ observed by \citet{heyer04} in M33. The galaxy M33 is 
very interesting. It is a nearly isolated, small disk galaxy with low
luminosity and low mass.  It has total mass of $M_{\rm tot} \sim 10^{11}\ \rm
M_{\odot}$ and a gas mass of $\rm M_{\rm gas} \sim 8.0 \times 10^{9}\ \rm
M_{\odot}$ \citep{heyer04}. It is molecule-poor and sub-critical, with gas
surface density is much smaller than the threshold surface density for star 
formation found by \citet{martin01}. However, it is actively forming stars
\citep{heyer04}. We do not have a model that exactly resembles M33, although a
close one might be model G100-1 in terms of mass. However, the gas velocity
dispersion of M33 is unknown, so we cannot make a direct comparison with our
G100-1 models. In Figure~\ref{fig_local_fit}, the low-$T$ model G100-1 has $N
\sim 1.4$, but we have not derived a value for its high-$T$ counterpart, as it
does not form stars at all in the first 3 Gyrs. Any stars that form in a disk
similar to this will likely form in spiral arms or other nonlinear density
perturbations that are not well characterized by an azimuthally averaged
stability analysis.  If these perturbations occur in the highest surface
density regions as might be expected, the local Schmidt law will have a very
high slope as observed. This speculation will need to be confirmed with models
reaching higher mass resolution in the future.  The details of the
feedback model and equation of state may also begin to play a role in
this extreme case.

The averaged values of our derived local Schmidt laws do agree
well with the observations by \citet{wong02} of a number of other
nearby galaxies. The similarity between the global and local Schmidt laws
suggests a common origin of the correlation between $\Sigma_{\rm {SFR}}$ and
$\Sigma_{\rm {gas}}$ in gravitational instability. 

\section{STAR FORMATION EFFICIENCY}
\label{sec_sfe}
 
The SFE is poorly understood, because it is difficult in both observations and
simulations to determine the timescale for gas removal and the gaseous and
stellar mass within the star formation region. On the molecular cloud scale,
observations of several nearby embedded clusters with mass $M <1000 \rm
M_{\odot}$ indicate that the SFEs range from approximately 10--30\%
\citep{lada03}. However, it is thought that field stars form with SFE
of only 1--5\% in giant molecular clouds (e.g., \citealt{duerr82}),
while the formation of a bound stellar cluster requires a local SFE
$\gtrsim$ 20--50\% (e.g., \citealt{wilking83, elmegreen97b}). An analytical
model including outflows by \citet{matzner00} suggests that the efficiency of
cluster formation is in the range of 30--50\%, and that of single star formation
could be anywhere in the range 25--70\%.

In the analysis of our simulations presented here, we convert the mass
of the sink particles into stars using a fixed local SFE $\epsilon_{\ell} =
30$\%, consistent with both the observations and theoretical predictions 
mentioned above.  This local efficiency is different from the global
star formation efficiency in galaxies $\epsilon_{g} \leq \epsilon_{\ell}$,
which measures the fraction of the {\em total} gas turned into stars. On a
galactic scale, the star formation efficiency appears to be associated with
the fraction of molecular gas (e.g., \citealt{rownd99}). The global SFE 
has had values derived from observations over a wide range, depending on the
gas distribution and the molecular gas fraction \citep{k98}. For example, in
normal galaxies $\epsilon_{g} \simeq 2$--10\%, while in starburst galaxies
$\epsilon_{g} = 10$--50\%, with a median value of 30\%.  One factor that 
appears to contribute to the differences in $\epsilon_{g}$ is
the gas content.  The global SFE is generally averaged over all gas
components, but since star formation correlates tightly with the local
gravitational instability one expects higher global SFE in more
unstable galaxies. In fact, as pointed out by \citet{wong02}, most
normal galaxies in the sample of \citet{k98} are molecule-poor
galaxies, which seem to have high stability and low SFE, while
molecule-rich starburst galaxies appear to be unstable, forming stars
with high efficiency.

The variation of the normalization of the local Schmidt laws in
\S~\ref{sec_local} also suggests that the global SFE varies from
galaxy to galaxy. To quantitatively measure the SFE in our models, we
apply the common definition of the global SFE, 
\begin{equation}
\epsilon_{g} = M_{*}/(M_{*}+ M_{\rm gas}) = M_*/M_{\rm init} \label{eglobal}
\end{equation} 
over a period of $10^8$ years, an average timescale for star formation in galaxy.
In this equation, $M_{*}= \epsilon_{\ell} M_{\rm sink}$ is the mass of newly
formed stars,  and $M_{\rm gas}$ includes both the remaining mass of the
sink particles and the SPH particles, so that $M_{*}+M_{\rm gas} = M_{\rm
  init}$ is the total mass of the initial gas. 
 
\begin{figure}
\begin{center}
\includegraphics{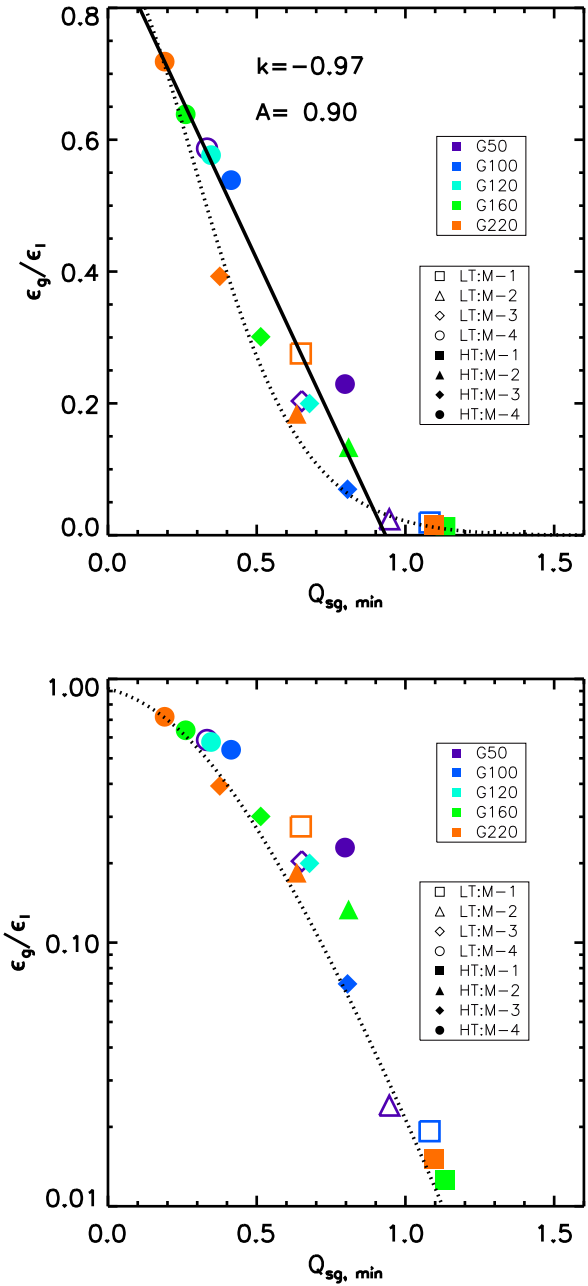}
\caption[Global Star Formation Efficiency]{\label{fig_effg} The global SFE
normalized by the local one $\epsilon_{\rm g}/\epsilon_{\ell}$ versus the
minimum initial gravitational instability parameter $Q_{sg,{\rm min}}$ in linear
({\em a}) and log space ({\em b}). The legends are the same 
as in Figure~\ref{fig_global}. The black solid line is the
least-absolute-deviation fit to the data, while the dotted line is the
function given in equation (\ref{eq_eg}).}  
\end{center} 
\end{figure}

Figure~\ref{fig_effg}({\em a}) and ({\em b}) show the relation between the
minimum values of the initial $Q$ parameter $Q_{sg, {\rm min}}$ and the global
SFE normalized by the local SFE $\epsilon_{\rm g}/\epsilon_{\ell}$.  The
time period is taken as the first 100 Myr after star formation starts.
If we take $\epsilon_{\ell}$ as a constant for all models, it appears
that $\epsilon_{g}$ declines as $Q_{sg,{\rm min}}$ increases.
Therefore, $\epsilon_{g}$ is high in less stable galaxies with high
mass or high gas fraction. A least-absolute-deviation fit of the data
gives a linear fit of $\epsilon_{g}/\epsilon_{\ell} = 0.9 -0.97\ 
Q_{sg,{\rm min}}$. 
This fit is good for values of $Q_{sg,{\rm min}} \le 1$.
For more stable galaxies, with larger values of $Q_{sg,{\rm min}}$, the
SFE remains finite, deviating from the linear fit.

Using the empirical relations we have derived from our models earlier
in the paper, we can derive a better analytic expression for $\epsilon_{g}$. 
Equation~(\ref{eglobal}) can be combined with equation~(\ref{eq_mas})
in \S~\ref{subsec_sfr} to yield an equation for
the global SFE 

\begin{equation} 
\epsilon_{g}  =  M_0\left[1 -
  \exp\left({-t}/{\tau_{\rm SF}}\right)\right]
\end{equation}

We evaluate this at $t = 100$~Myr, taking the definitions of $M_0$ and
$\tau_{\rm SF}$ derived from equations (\ref{eq_m0}) and
(\ref{eq_tau}).  Normalizing by the local SFE, we find

\begin{align}
\label{eq_eg}
\epsilon_g / \epsilon_{\ell} = 0.96\times \left[1 -
  2.88\exp\left({-1.7}/{Q_{sg, {\rm min}}}\right)\right] \notag \\
\times \left[1 - \exp \left(-2.9e^{-Q_{sg, {\rm min}}/0.24}\right)\right]  
\end{align}
The function given by equation~(\ref{eq_eg}) is shown in
Figure~\ref{fig_effg}. For $Q_{sg,{\rm min}} \le 1.0$ it is well
approximated by the much simpler linear function
\begin{equation}
\epsilon_{g}/\epsilon_{\ell} \simeq 0.9-Q_{sg,{\rm min}}
\end{equation} 
as shown in Figure~\ref{fig_effg}({\em a}). At larger values of
$Q_{sg,{\rm min}}$, the exact function predicts the 
SFE in our models excellently as shown in Figure~\ref{fig_effg}({\em
  b}).  Observational verification of this behavior is vital.

\section{ASSUMPTIONS AND LIMITATIONS OF THE MODELS}
\label{sec_dis}

\subsection{Isothermal Equation of State}

One of the two central assumptions in our model is the use of an isothermal
equation of state to represent a constant velocity dispersion. This is of
course a simplification, as the interstellar medium in reality has a broad
range of temperatures 10~K$< T < 10^7$~K. However, neutral gas velocity
dispersions in normal spiral galaxies cover a far more limited range,
as reviewed by \citet{k98,elmegreen04,scalo04} and \citet{dib05}. The
characteristic $\sigma$ increases when the averaged $\Sigma_{\rm SFR}$ of a
galaxy reaches tens of solar masses per year, but the normal galaxies in the
sample with reliable measurements lie in the range $\sim$7--13~km~s$^{-1}$
(e.g., \citealt{elmegreen84, meurer96, vanzee97, stil02, hippelein03}).  

At least two mechanisms appear viable for maintaining roughly constant
velocity dispersion for the bulk of the gas in a galactic disk,
supernova feedback and magnetorotational instability \citep{mk04}.
Three-dimensional simulations in a periodic box with parameters characteristic 
of the outer parts of galactic disks by \citet{dib05} show that supernova
driving leads to constant velocity dispersions of $\sigma \sim$ 6~km~s$^{-1}$
for the total gas and $\sigma_{\rm H_{\rm I}} \sim$ 3~km~s$^{-1}$ for the
H$_{\rm I}$ gas, independent of the supernova rate. Simulations of the
feedback effects across whole galactic disks do suggest that the inner parts
have slightly higher velocity dispersions (e.g. \citealt{thacker00}), though
within the range that we consider. The magnetorotational instability in
galactic disks was suggested by \citet{sellwood99} to maintain the observed
velocity dispersion, a suggestion that has since been substantiated by both local
\citep{piontek05} and global \citep{dziour04} numerical models.  This
may act even in regions with little or no active star formation.

Recently, \citet{robertson04} presented simulations of galactic disks
and claimed that an isothermal equation of state leads to a collapsed
disk as the gas fragments into clumps that fall to the galactic center due to
dynamical friction. However, similar behavior is seen in models by
\citet{immeli04} who did not use an isothermal equation of state, but
also ran at resolutions not satisfying the Jeans criterion
\citep{bate97,truelove97}. On the other hand, using essentially the same code 
and galaxy model as \citet{robertson04}, but with higher resolution satisfying
the Jeans criterion, we do not see this collapse. Insufficient resolution that
fails to resolve the Jeans mass leads to spurious, artificial fragmentation
and thus collapse.

Simulations by \citet{governato04} suggest that some long-standing
problems in galaxy formation such as the compact disk and lack of
angular momentum may well be due to insufficient resolution or
violation of numerical criteria. Our results lead us to agree that the
isothermal equation of state is not the cause of the compact disk
problem, but rather inadequate numerical resolution.

Our assumption of an isothermal equation of state does, of course,
rule out the treatment by our model of phenomena such as galactic
winds associated with the hot phase of the interstellar medium
(although the venting of supernova energy vertically may help maintain
the isothermal behavior of the gas in the plane).  The strong
starbursts produced in some of our galaxy models will certainly cause
strong galactic winds.  It remains unclear whether even strong
starbursts can remove substantial amounts of gas, though.  Certainly
they cannot in small galaxies \citep{mf99}, and larger galaxies would
seem more resistant to stripping in starbursts than smaller ones.
However, galactic winds will certainly influence the surroundings of
starburst galaxies, as well as their observable properties.  These
effects should eventually be addressed in future simulations with more
comprehensive gas physics and a more realistic description of the
feedback from star formation.

\subsection{Sink Particles}
The use of sink particles enables us to directly identify high gas
density regions, measure gravitational collapse, and follow the dynamical
evolution of the system to a long time. We can therefore determine the star
formation morphologies and rates, and study the Schmidt
laws and star formation thresholds.   

However, one shortcoming of our sink particle implementation is that
we do not include gas recycling. Once the gas collapses into the
sinks, it remains locked up there. As discussed in Paper I, the bulk
of the gas that does not form stars will remain in the disk and
contribute to the next cycle of star formation. Also, the ejected
material from massive stars will return into the gas reservoir
for future star formation.  Another problem is accretion. In the
current model, sink particles accrete until the surrounding gas is
completely consumed. However, in real star clusters, the accretion
would be cut-off due to stellar radiation, and the clusters will actually
lose mass due to outflow and tidal stripping.

These shortcomings of our sink particle technique may contribute to two
limitations of our models: first, the decline of star formation rate over time
due to consumption of the gas, as seen in Figure~\ref{fig_sfr}; second, the
variation of SFR over time in the simulated global Schmidt laws in
Figure~\ref{fig_par_t}. Nevertheless, as we have demonstrated in the previous
sections, our models are valid within one gas consumption time $\tau_{\rm
SF}$, and are sufficient to investigate the dominant physics that controls
gravitational collapse and star formation within that period.

\subsection{Initial Conditions of Galaxies}
Many nearby galaxies appear to be gas-poor and stable ($Q_{sg} > 1$).
However, their progenitors at high redshift were gas rich, so the bulk
of star formation should have taken place early on. In order to test
this, we vary the gas fraction (in terms of total disk mass) in the
models. We also vary the total galaxy mass and thus the rotational
velocity. These result in different initial stability curves.
Massive, or gas-rich galaxies have low values of the $Q$ parameters,
so they are unstable, forming stars quickly and efficiently.

There are no observations yet that directly measure the $Q$ values in
starburst galaxies. However, indirectly, observations by
\citet{dalcanton04} show that dust lanes, which trace star formation,
only form in unstable regions.  Moreover, observations of color
gradients in disk galaxies by \citet{macarthur04} show that massive
galaxies form stars earlier and with higher efficiency. Both of these
observations are naturally explained by our models.

The Toomre $Q$ parameter for gas $Q_g$ differs from that for a
combination of stars and gas $Q_{sg}$ in some of our model galaxies.
This leads to slightly different results in Figure~\ref{fig_sfr_fit}
and Figure~\ref{fig_sfr_q} where we compare $Q_{sg,{\rm min}}$ and
$Q_{g,{\rm min}}$. However, we believe $Q_{sg}$ is a better measure of
gravitational instability in the disk, as it takes into account both the
collisionless and the collisional components, and the interaction between
them. We note that it is a simplified approach to quantify the instability of
the entire disk with just a number $Q_{sg,{\rm min}}$, as $Q$ has a radial
distribution, evolves with time, and is an azimuthally averaged
quantity, but nevertheless, we find interesting regularities by making
this approximation.

\section{SUMMARY}
\label{sec_sum}

We have simulated gravitational instability in galaxies with sufficient
resolution to resolve collapse to molecular cloud pressures in models
of a wide range of disk galaxies with different total mass, gas
fraction, and initial gravitational instability. Our calculations are
based on two approximations: the gas of the galactic disk has an
isothermal equation of state, representing a roughly constant gas
velocity dispersion; and sink particles are used to follow
gravitationally collapsed gas, which we assume to form both stars and
molecular gas. With these approximations, we have derived star
formation histories; radial profiles of the surface density of
molecular and atomic gas and SFR; both the global and local Schmidt
laws for star formation in galaxies; and the star formation
efficiency.

The star formation histories of our models show the exponential
dependence on time given by equation~\ref{eq_sfr} in agreement with,
for example, the interpretation of galactic color gradients by
\citet{macarthur04}.  The radial profiles of atomic and molecular gas
qualitatively agree with those observed in nearby galaxies, with
surface density of molecular gas peaking centrally at values much
above that of the atomic gas \citep[e.g.][]{wong02}.  The radial
profile of the surface density of SFR correlates linearly with that of
the molecular gas, agreeing with the observations of \citet{gao04b}.

Our models quantitatively reproduce the observed global Schmidt law
\citep{k98}---the correlation between the surface density of star
formation rate $\Sigma_{\rm {SFR}}$ and the gas surface density
$\Sigma_{\rm {gas}}$---in both the slope and normalization over a wide
range of gas surface densities (eq.~\ref{gsl}).  We show that $\Sigma_{\rm
  SFR}$ is strongly correlated with the gravitational instability of galaxies
$\Sigma_{\rm SFR} \propto \left[Q_{sg,{\rm min}}(\tau_{\rm SF})\right]^{-1.54
  \pm 0.23}$, where $Q_{sg,{\rm min}}(\tau_{\rm SF})$ is the local instability
parameter at time $t=\tau_{\rm SF}$ (see eq.~\ref{eq_sfr_q_fit}). This
correlation naturally leads to the Schmidt law.

On the other hand, our models do not reproduce the correlation $\Sigma_{\rm
SFR} \sim \Sigma_{\rm gas} \Omega$ derived from kinematical models
\citep{k98}. However, they may agree better with the dependence of the normal
galaxies on this quantity, as suggested by \citet{boissier03}. The discrepancy
may be caused by the lack of extreme starburst galaxies such as galaxy mergers
in our set of models.

The local Schmidt laws of individual galaxies clearly show evidence of star
formation thresholds above a critical surface density. The threshold surface
density varies with galaxy, and appears to be determined by the gravitational
stability of the disk. The derived threshold parameters for our stable models  
cover the range of values seen in observations of normal galaxies. The local
Schmidt laws have significant variations in both slope and normalization, but 
also cover the observational ranges reported by \citet{wong02},
\citet{boissier03} and \citet{heyer04}. The average normalization and slope of
the local power-laws are very close to those of the global Schmidt law. 

Our models show that the global star formation efficiency (SFE)
$\epsilon_{g}$ can be quantitatively predicted by the gravitational
instability of the disk.  We have used a fixed local SFE
$\epsilon_{\ell} = 30\%$ to convert the mass of the sink particles to
stars in our analysis. This is a reasonable assumption for the SFE in
dense, high pressure molecular clouds. The global SFE of a galaxy then
can be shown to depend quantitatively on a nonlinear function
(eq.~\ref{eq_eg}) of the minimum Toomre parameter $Q_{sg,{\rm min}}$
for stars and gas that can be approximated for
$Q_{sg,{\rm min}} \le 1.0$ with the linear correlation $\epsilon_{\rm
  g}/\epsilon_{\ell} \propto 0.9 -Q_{sg,{\rm min}}$.  More unstable
galaxies have higher SFE.  Massive, or gas-rich galaxies in our suite
of models are unstable, forming stars quickly with high efficiency.
They represent starburst galaxies.  Small, or gas-poor galaxies are
rather stable, forming stars slowly with low efficiency, corresponding
to quiescent, normal galaxies.

\acknowledgments We thank V. Springel for making both GADGET and his
galaxy initial condition generator available, as well as for useful
discussions, and A.-K. Jappsen for participating in the implementation of
sink particles in GADGET. We are grateful of F. Adams, G. Bryan, J. Dalcanton,
B. Elmegreen, D. Helfand, R. Kennicutt, J. Lee, C. Martin, R. McCray,
T. Quinn, E. Quataert, M. Shara, and J. van Gorkom for very useful
discussions. We also thank the referee, Dr. C. Struck for valuable comments
that help to improve this manuscript. This work was
supported by the NSF under grants AST99-85392 and AST03-07854, by NASA under grant
NAG5-13028, and by the Emmy Noether Program of the DFG under grant KL1358/1.
Computations were performed at the Pittsburgh Supercomputer Center supported
by the NSF, on the Parallel Computing Facility of the AMNH, and on an
Ultrasparc III cluster generously donated by Sun Microsystems.

\clearpage

\begin{deluxetable}{lllllllllll}
\tablecolumns{11}
\tablecaption{Galaxy Models and Numerical Parameters \label{tab1}}
\tablehead{\colhead{Model\tablenotemark{a}} & 
\colhead{$f_{\rm g}$\tablenotemark{b}} & 
\colhead{$R_{\rm d}$\tablenotemark{c}} & 
\colhead{$Q_{sg}$(LT)\tablenotemark{d}} & 
\colhead{$Q_{sg}$(HT)\tablenotemark{e}} & 
\colhead{$N_{\rm tot}$\tablenotemark{f}} & 
\colhead{$h_{\rm g}$\tablenotemark{g}} & 
\colhead{$m_{\rm g}$\tablenotemark{h}} &
\colhead{$\tau_{\rm SF}$(LT)\tablenotemark{i}} &
\colhead{$\tau_{\rm SF}$(HT)\tablenotemark{j}}}  
\startdata
G50-1  & 0.2  &  1.41  &  1.22  &  1.45  & 1.0  & 10 &  0.08     & 4.59 & \nodata \\
G50-2  & 0.5  &  1.41  &  0.94  &  1.53  & 1.0  & 10 &  0.21     & 1.28 & \nodata \\
G50-3  & 0.9  &  1.41  &  0.65  &  1.52  & 1.0  & 10 &  0.37     & 0.45 & \nodata \\
G50-4  & 0.9  &  1.07  &  0.33  &  0.82  & 1.0  & 10 &  0.75     & 0.15 & 0.53 \\
G100-1 & 0.2  &  2.81  &  1.08  &  1.27  & 6.4  & 7  &  0.10  & 2.66 & \nodata  \\
G100-2 & 0.5  &  2.81  &  \nodata  &  1.07  & 1.0  & 10 &  1.65  & \nodata & \nodata \\
G100-3 & 0.9  &  2.81  &  \nodata  &  0.82  & 1.0  & 10 &  2.97  & \nodata & 1.92    \\
G100-4 & 0.9  &  2.14  &  \nodata  &  0.42  & 1.0  & 20 &  5.94  & \nodata & 0.15    \\
G120-3 & 0.9  &  3.38 &  \nodata  &  0.68  & 1.0  & 20 &  5.17   & \nodata & 0.46    \\
G120-4 & 0.9  &  2.57 &  \nodata  &  0.35  & 1.0  & 30 &  10.3   & \nodata & 0.16    \\
G160-1 & 0.2  &  4.51 &  \nodata  &  1.34  & 1.0  & 20 &  2.72   & \nodata & 3.1\tablenotemark{k}    \\
G160-2 & 0.5  &  4.51 &  \nodata  &  0.89  & 1.0  & 20 &  6.80   & \nodata & 0.58    \\
G160-3 & 0.9  &  4.51 &  \nodata  &  0.52  & 1.0  & 30 &  12.2   & \nodata & 0.30    \\ 
G160-4 & 0.9  &  3.42 &  \nodata  &  0.26  & 1.5  & 40 &  16.3   & \nodata & 0.11    \\
G220-1 & 0.2  &  6.20 &  0.65  & 1.11  & 6.4  & 15 &  1.11   & 0.28   & 3.0\tablenotemark{k} \\
G220-2 & 0.5  &  6.20 &  \nodata  &  0.66  & 1.2  & 30 &  14.8   & \nodata & 0.39  \\
G220-3 & 0.9  &  6.20 &  \nodata  &  0.38  & 2.0  & 40 &  15.9   & \nodata & 0.25  \\
G220-4 & 0.9  &  4.71 &  \nodata  &  0.19  & 4.0  & 40 &  16.0   & \nodata & 0.096 \\
\enddata \tablenotetext{a}{First number is rotational velocity in km
  s$^{-1}$ at the virial radius, the second number indicates
  sub-model. Sub-models have varying fractions  $m_{\rm d}$ of total
  halo mass in their disks, and given values of $f_{\rm g}$. Sub-models 1 -- 3
  have $m_{\rm d}=0.05$, while sub-model 4 has $m_{\rm d}=0.1$.}
\tablenotetext{b}{Fraction of disk mass in gas.}
\tablenotetext{c}{Stellar disk radial exponential scale length in kpc}
\tablenotetext{d}{Minimum initial value of $Q_{sg}(R)$ for low-$T$ models.}  
\tablenotetext{e}{Minimum initial value of $Q_{sg}(R)$ for high-$T$ models.} 
\tablenotetext{f}{Total particle number in units of $10^6$}
\tablenotetext{g}{Gravitational softening length of gas in pc.}
\tablenotetext{h}{Gas particle mass in units of $10^4\ \rm M_{\odot}$.}
\tablenotetext{i}{Star formation timescale in Gyr  of low-$T$ model
  (from Paper I).}
\tablenotetext{j}{Star formation timescale in Gyr  of high-$T$ model
  (from Paper I).}
\tablenotetext{k}{Maximum simulation timestep
  instead of the star formation timescale $\tau_{\rm SF}$.} 
\end{deluxetable}

\end{document}